\documentclass[10pt,onecolumn,aps,pra,floats,superscriptaddress]{revtex4-1}

	\usepackage{amsmath,amssymb,amsthm} 
	\usepackage{bm} 					

	\usepackage{graphicx} 				
		\usepackage{color}				
		\usepackage{transparent}		

	\usepackage{microtype}			
	\usepackage{upgreek}

	\let\REALPART\Re
	\renewcommand{\Re}[1]{\ensuremath{\REALPART\left\{ #1 \right\} }}
	\let\IMAGPART\Im
	\renewcommand{\Im}[1]{\ensuremath{\IMAGPART\left\{ #1 \right\} }}

	\catcode`_=\active
	\newcommand_[1]{\ensuremath{\sb{\mathrm{#1}}}}

	\newcommand{\de}[2]{\ensuremath{\frac{\textup{d} #1}{\textup{d} #2}}}   

	\newcommand{\intfin}[3]{\ensuremath{\int\sb{#2}^{#3}} \textup{d}#1 \;} 
	\newcommand{\intinf}[1]{\ensuremath{\int\sb{-\infty}^{\infty}} \textup{d}#1 \;} 

	\renewcommand{\mod}[1]{\ensuremath{\left| #1 \right|}} 
	\newcommand{\td}{\ensuremath{\left(t\right)}} 
	\newcommand{\Tr}[1]{\ensuremath{\textup{Tr}\left\{ #1 \right\}}} 
	\newcommand{\trans}[1]{\ensuremath{ \left.#1\right.^{\mathrm{T}}}} 

	\newcommand{\Br}[1]{\ensuremath{\left( #1 \right)}} 
	\newcommand{\Sq}[1]{\ensuremath{\left[ #1 \right]}} 
	\newcommand{\Cu}[1]{\ensuremath{\left\{ #1 \right\}}} 
	\newcommand{\ghost}[1]{\ensuremath{\left. #1 \right.}} 
	\newcommand{\upBr}[2]{\ensuremath{\ghost{#1^{\Br{#2}}}} } 

	\newcommand{\ket}[1]{\ensuremath{ \left|  #1 \right> }}         
	\newcommand{\expect}[1]{\ensuremath{\left< #1 \right>}}
	\newcommand{\comm}[2]{\ensuremath{\left[ #1,#2 \right]}}
	\newcommand{\D}[1]{\ensuremath{\hat{D}\left( #1 \right)}} 
	\newcommand{\n}{\ensuremath{\bar{n}}} 
	\newcommand{\baop}[1]{\ensuremath{\hat{#1}}}   
	\newcommand{\bcop}[1]{\ensuremath{\hat{#1}^\dagger}}   

	\newcommand{\I}{\mathrm{i}} 				
	\newcommand{\e}{\mathrm{e}} 				

	\newcommand{\R}{{\mathbb R }} 

	\newcommand{\half}{\ensuremath{\frac{1}{2}}}   

	\newcommand{\ie}{\textit{i.e.}} 
	\newcommand{\eg}{\textit{e.g.}} 
	\newcommand{\viz}{\textit{viz}.} 
	\newcommand{\cf}{\textit{cf}.} 
	\newcommand{\ala}{\textit{\`{a} la}} 
	\newcommand{\naive}{na\"\i{}ve} 

	\newcommand{\Schrod}{Schr\"{o}dinger} 
	\newcommand{\BCH}{Baker--Campbell--Haussdorf}

\begin{document}

\title{A Quantum Optomechanical Interface Beyond the Resolved Sideband Limit}

\author{James S. Bennett}
\email{Corresponding author: james.bennett2@uqconnect.edu.au}
	\affiliation{Australian Research Council Centre of Excellence for Engineered Quantum Systems (EQuS), School of Mathematics and Physics, The University of Queensland, St Lucia, QLD 4072, Australia}
\author{Kiran Khosla}
	\affiliation{Australian Research Council Centre of Excellence for Engineered Quantum Systems (EQuS), School of Mathematics and Physics, The University of Queensland, St Lucia, QLD 4072, Australia}
\author{Lars S. Madsen}
	\affiliation{Australian Research Council Centre of Excellence for Engineered Quantum Systems (EQuS), School of Mathematics and Physics, The University of Queensland, St Lucia, QLD 4072, Australia}
\author{Michael R. Vanner}
	\affiliation{Australian Research Council Centre of Excellence for Engineered Quantum Systems (EQuS), School of Mathematics and Physics, The University of Queensland, St Lucia, QLD 4072, Australia}
	\affiliation{Clarendon Laboratory, University of Oxford, OX1 3PU, United Kingdom}
\author{Halina Rubinsztein-Dunlop}
	\affiliation{Australian Research Council Centre of Excellence for Engineered Quantum Systems (EQuS), School of Mathematics and Physics, The University of Queensland, St Lucia, QLD 4072, Australia}
\author{Warwick P. Bowen}
	\affiliation{Australian Research Council Centre of Excellence for Engineered Quantum Systems (EQuS), School of Mathematics and Physics, The University of Queensland, St Lucia, QLD 4072, Australia}

\begin{abstract}
Mechanical oscillators which respond to radiation pressure are a promising means of transferring quantum information between light and matter. Optical--mechanical state swaps are a key operation in this setting. Existing proposals for optomechanical state swap interfaces are only effective in the resolved sideband limit. Here, we show that it is possible to fully and deterministically exchange mechanical and optical states outside of this limit, in the common case that the cavity linewidth is larger than the mechanical resonance frequency. This high-bandwidth interface opens up a significantly larger region of optomechanical parameter space, allowing generation of non-classical motional states of high-quality, low-frequency mechanical oscillators.
\end{abstract}

\date{\today}

\maketitle

\section{Introduction} \label{Sec:Introduction}

Quantum interfaces are anticipated to form a crucial component of future quantum information networks due to their ability to transfer quantum information between `flying' carriers---photons---and `stationary' quantum media \cite{Kimble2008,Wallquist2009}. Numerous physical settings lend themselves to creating such interfaces, including both warm and cold neutral atoms \cite{Hammerer2004,Hammerer2010a}, quantum dots \cite{Peter2005,Montinaro2014}, nitrogen--vacancy (and other) colour centres \cite{Putz2014,Grezes2014} and trapped ions \cite{Dantan2009,Stute2012}.

Cavity optomechanical systems are a promising platform for novel quantum interfaces because they may be used to couple light to the `stationary' media listed above, and because existing mechanical oscillators have extremely long decay times ($\sim 30$~s, \eg{} \cite{Brawley2012}). In these systems a mechanical degree of freedom with (angular) frequency $\omega_{M}$ is parametrically coupled to an optical cavity with linewidth $\kappa$. These devices have entered the quantum regime for the first time in recent years  (\eg{} \cite{Chen2013, OConnell2010, Teufel2011, Chan2011}) and they are set to find technological and research applications in a variety of capacities. For instance, a mechanical oscillator which is subject to radiation pressure arising from microwave and optical modes simultaneously may be used to couple superconducting circuits to optics \cite{Bochmann2013,Bagci2014,Tian2015,Xia2014,Wang2012,Andrews2014}; the capability to do this at the quantum level would permit construction of quantum information networks that utilise the advantages of both low-loss optical transmission and the exquisite quantum control of superconducting circuits. There are also a wealth of proposals for employing cavity optomechanical coupling to prepare massive mechanical oscillators in non-classical states, which could allow sensitive probing of the quantum-to-classical transition \cite{Chen2013} or quantum-enhanced metrology (\eg{} mass sensing \cite{Liu2013}, accelerometry \cite{Krause2012}).

In many of these applications it is expedient to employ a low-frequency mechanical oscillator (\eg{} \cite{Milburn2011}); for instance, devices with material-limited damping rates $\Gamma$ obey the scaling $\Gamma \propto \omega_{M}^{2}$ (due to the Akhiezer effect \cite{Chandorkar2008}). Furthermore, large optomechanical coupling rates require small cavity volumes, and the optical linewidth typically scales inversely with its length, so that $\kappa$ is often unavoidably larger than $\omega_{M}$ in micro- and nano-optomechanical devices. This high-bandwidth operating regime, $\kappa \gg \omega_{M}$, is known as the \textit{unresolved sideband limit} or \textit{bad-cavity limit}, despite the fact that the cavities involved often have large $Q$ factors.

Current technologies make it possible to cool a mechanical oscillator to near its ground state in the unresolved sideband limit using optical measurement-based feedback cooling (\eg{} \cite{Poggio2007}), or to perform steady-state cooling using hybrid quantum systems \cite{Bennett2014,Joeckel2015,Chen2015}, dissipative optomechanics \cite{Elste2009}, optomechanically-induced transparency \cite{Ojanen2014}, and related schemes (\eg{} \cite{Genes2009,Gu2013}). Very rapid ground state cooling of the mechanical oscillator is also enabled by varying the optical drive's amplitude and phase dynamically \cite{Vanner2011, Khosla2013, Vanner2013, Machnes2012,Wang2011b}. Alternatively, one may exploit the large cavity bandwidth in the bad-cavity limit to engineer quantum non-demolition (QND) interactions between short optical pulses and a mechanical oscillator (\cf{} \eqref{Eqns:QND}) \cite{Vanner2011}. QND interactions are the basis of proposed quantum `upload' interfaces, which write the state of one input mode onto one output mode \cite{Filip2008,Marek2010,Khalili2010}. Further applications of QND interactions include generation of mechanical geometric phases \cite{Khosla2013}, and cooling by measurement and state tomography \cite{Vanner2013}.

Despite this---and the success of optomechanical interfaces in the `resolved-sideband' ($\kappa \ll \omega_{M}$), regime (\eg{} \cite{Marquardt2007,Teufel2011,Chan2011,Palomaki2013,Vitali2007,Hofer2011})---there are currently no quantum optical--to--mechanical interfaces which permit a state swap in the unresolved sideband limit. Teleportation schemes \cite{Mancini2003} and the aforementioned quantum `upload' interfaces \cite{Filip2008,Marek2010}, which transfer an input state from one degree of freedom to another, are incapable of directly exchanging two input states.

Here we present a \textit{bidirectional} optomechanical interface which can perform a quantum state swap and other operations in the unresolved sideband limit. Our interface is completely deterministic and does not rely on measurement or any form of conditioning or post-selection. Furthermore, the proposed protocol only requires classical open-loop control, which may be performed with very low added noise (\ie{} below the back-action noise) \cite{Harris2012a}.

We predict that the interface is capable of performing a number of important operations, including mechanical ground state cooling, squeezing, and exploiting non-Gaussian optical inputs to create Wigner-negative mechanical states, including low-amplitude \Schrod{} cat (`kitten') states.

\section{Pulsed QND Interaction} \label{Sec:PulsedQND}

We now show that QND interactions may be performed by pulsed optomechanics, following \cite{Vanner2011}. Consider a mechanical oscillator with annihilation operator $\baop{b}$ and angular frequency $\omega_{M}$ parametrically coupled to an optical cavity. As the (dimensionless) position of the oscillator $X_{M} = \bcop{b} + \baop{b}$ (momentum $P_{M} = \I (\bcop{b} - \baop{b})$) fluctuates the resonance frequency of the cavity also shifts. Thus the optomechanical (dispersive) coupling Hamiltonian takes the form \cite{Milburn2011, Aspelmeyer2013review}
\[
	\hat{H} = \hbar g_{0} X_{M}\bcop{a}\baop{a},
\]
where $\baop{a}$ is the cavity annihilation operator and the optomechanical coupling rate is $g_{0}$. This is understood as the frequency shift of the cavity per zero-point displacement of the oscillator. For a detailed discussion of the optomechanical Hamiltonian the reader is referred to \cite{Milburn2011}.

We assume that $g_{0} \ll \omega_{M}$, as is usually the case (\eg{} review of parameters given by \cite{Anetsberger2011}). Then if a bright pump beam is applied to the cavity the interaction between the cavity field and mechanical fluctuations is well described by the effective linearised Hamiltonian \cite{Vanner2011}
\[
	\hat{H} = \hbar g_{0}\mod{\alpha\td} X_{M} \bar{X}_{L}\td + \hbar g_{0} X_{M} \mod{\alpha\td}^{2}.
\]
Here, $\alpha\td$ is the mean intracavity amplitude induced by the pump, and the optical fluctuation operator $\bar{X}_{L}\td = \bcop{a}+\baop{a}-2\alpha\td$ describes amplitude fluctuations of the field (this expression is valid for real, positive $\alpha$; for the general case see Appendix \ref{App:FluctuationsDef}). These quantities are written with explicit time dependence to show that they depend on the pulse used to illuminate the cavity. The phase fluctuations of the optical field $\bar{P}_{L}\td$ do not appear in the interaction Hamiltonian. Both fluctuation operators obey $\expect{\bar{X}_{L}\td} = \expect{\bar{P}_{L}\td} = 0$ for all times.

We now consider applying pulses of light to the cavity. These are described by a mean amplitude $\alpha_{in}\td$, normalised such that the average number of photons in the input pulse envelope is
\[
	N = \intinf{t} \; \mod{\alpha_{in}\td}^{2} \gg 1,
\]
and the instantaneous amplitude and phase noise operators, $\bar{X}_{L,in}\td$ and $\bar{P}_{L,in}\td$. The latter are defined by analogy with the intracavity field.
The optical field exiting the cavity is connected to the intracavity mode and input field through the input--output relation. This takes the form 
\begin{eqnarray}
	\bar{X}_{L,out}\td = \sqrt{\kappa}\bar{X}_{L}\td - \bar{X}_{L,in}\td , 
	\label{Eqn:InputOutput}
\end{eqnarray}
with a similar expression holding for the phase fluctuations \cite{WallsMilburn2008}.

If the bandwidth of the pulse is large compared to both the mechanical frequency and the thermal heating rate $\Gamma\times\Br{\n_{B}+1/2}$ we may neglect the damped dynamics of the oscillator over the duration of the interaction; furthermore, the large cavity linewidth (compared to the pulse bandwidth) allows the optical fields to be adiabatically eliminated. Together, these approximations yield effective Langevin equations of the form
\begin{subequations}
\label{Eqns:EoM}
	\begin{eqnarray}
		\dot{X}_{M} & = & 0 \label{Eqn:EoMX}\\
		\dot{P}_{M} & = & -\frac{8g_{0}\mod{\alpha_{in}\td}}{\kappa}\Br{\bar{X}_{L,in}\td+\mod{\alpha_{in}\td}} \label{Eqn:EoMP}\\
		\bar{X}_{L,out}\td & = & \bar{X}_{L,in}\td \label{Eqn:EoMOX}, \\
		\bar{P}_{L,out}\td & = & \bar{P}_{L,in}\td-\frac{8g_{0}\mod{\alpha_{in}\td}}{\kappa}X_{M} \label{Eqn:EoMOP},
	\end{eqnarray}
\end{subequations}
in which we have used \eqref{Eqn:InputOutput} to obtain \eqref{Eqn:EoMOX} \& \eqref{Eqn:EoMOP}.

Integrating \eqref{Eqn:EoMX} and \eqref{Eqn:EoMP} over the duration of the pulse and employing the fact that the pulse envelope changes rapidly compared to $\omega_{M}^{-1}$ allows us to obtain the following transformation between the initial and final mechanical quadratures (final quadratures are primed).
\begin{eqnarray}
	X_{M}^{\prime} & = & X_{M}, \nonumber\\
	P_{M}^{\prime} & = & P_{M} + \chi \sqrt{N} + \chi \intinf{t} \; \frac{\mod{\alpha_{in}\td}}{\sqrt{N}}\bar{X}_{L,in}\td, \label{Eqn:QNDMechTotal}
\end{eqnarray}
where the interaction strength is $\chi = -8 g_{0} \sqrt{N}/\kappa$. We see that the weighted optical amplitude fluctuations have been transferred onto the mechanical momentum $P_{M}^{\prime}$, along with a classical momentum kick $\chi\sqrt{N}$. It is possible to subtract off the classical kick by performing a displacement on the mechanical oscillator immediately after the interaction; this can be done using open-loop control with very little added noise (\eg{} \cite{Harris2012a}). We will therefore neglect this kick hereafter.

Motivated by the form of \eqref{Eqn:QNDMechTotal}, we factorise the pulse envelope into the form $\alpha_{in}\td = \sqrt{N}f\td$; thus the integral of $\mod{f\td}^{2}$ over the whole pulse is unity. Multiplying \eqref{Eqn:EoMOX} \& \eqref{Eqn:EoMOP} by $\mod{f\td}$ and integrating, we find
\begin{equation}
	X_{L}^{\prime} = X_{L},\; P_{L}^{\prime} = P_{L} + \chi X_{M},  \label{Eqn:QNDOptAlone}
\end{equation}
in which the collective pulse quadratures $X_{L}$ and $P_{L}$ ($\comm{X_{L}}{P_{L}} = \comm{X_{M}}{P_{M}} = 2\I$) have been defined by
\begin{eqnarray*}
	X_{L} & = & \intinf{t} \; \mod{f\td}\bar{X}_{L,in}\td \\
	P_{L} & = & \intinf{t} \; \mod{f\td}\bar{P}_{L,in}\td.
\end{eqnarray*}

Note that these quadratures are zero-mean by construction. Similar relations hold for $X_{L}^{\prime}$ and $P_{L}^{\prime}$, with the input fields replaced by $\bar{X}_{L,out}\td$ and $\bar{P}_{L,out}\td$ respectively.

The evolution described by \eqref{Eqn:QNDMechTotal} (including a mechanical displacement) and \eqref{Eqn:QNDOptAlone} is thus a standard QND interaction between the mechanical oscillator and the collective quadratures of the optical pulse;
\begin{subequations}
\label{Eqns:QND}
	\begin{eqnarray}
	X_{M}^{\prime} = X_{M}, & \; & P_{M}^{\prime} = P_{M} + \chi X_{L}, \label{Eqn:QNDMech}\\
	X_{L}^{\prime} = X_{L}, & \; & P_{L}^{\prime} = P_{L} + \chi X_{M}.  \label{Eqn:QNDOpt}
	\end{eqnarray}
\end{subequations}
Each oscillator acquires information about one quadrature of the other during this interaction.

\section{Basic Protocol} \label{Sec:Model}

A QND interaction correlates each oscillator with the other, but only one quadrature at a time; evidently some extra ingredients will be required to perform a complete state swap. We will first show how to use a sequence of QND interactions and local operations to realize a state swap under ideal conditions. Secondly, we will generalise our protocol to allow both state transfer and squeezing. In our approach the pulse enters and leaves the optomechanical cavity multiple times, with local operations on the pulse and mechanical resonator occurring between the optomechanical interactions. Figure~\ref{Fig:Schema} summarises the necessary steps, as presented below, graphically.

This approach is in the spirit of the proposal of \cite{Filip2008,Marek2010}; however, our protocol replaces measurement and feedback with a third QND interaction to achieve a two-way interface.

Firstly, consider the case in which $\chi = -1$ and there are no optical or mechanical decoherence mechanisms. From \eqref{Eqns:QND} it is clear that a single QND pulse correlates $P_{M}^{\prime}$ ($P_{L}^{\prime}$) with $X_{L}$ ($X_{M}$). In order to build the necessary correlations between the other pair of quadratures we must exchange $X_{M}^{\prime}$ ($X_{L}^{\prime}$) and $P_{M}^{\prime}$ ($P_{L}^{\prime}$). This may be achieved by local $\pi/2$ rotations on each mode.

For definiteness, we can consider the local rotations to be generated in the following manner. Delay the pulse for a quarter of a mechanical period before re-injecting it into the cavity to achieve a mechanical rotation of $\pi/2$. The optical amplitude and phase noise rotations may be generated by \textit{displacing} the optical state in phase space, as demonstrated in Fig.~\ref{Fig:Schema} (panel \textit{ii}), \eg{} by interfering the pulse with a very bright, temporally matched pulse on a highly asymmetric beamsplitter (\cf{} Fig.~\ref{Fig:Schema}, panel \textit{iii}).

If we apply these rotations and then allow the pulse and oscillator to interact for a second time the quadratures have been transformed according to
\begin{subequations}
\label{Eqns:Pulse2}
	\begin{eqnarray}
	X_{M}^{\prime} = P_{M} - X_{L}, & \; & P_{M}^{\prime} = -P_{L}, \label{Eqn:Pulse2Mech}\\
	X_{L}^{\prime} = P_{L}-X_{M}, & \; & P_{L}^{\prime} = -P_{M}.  \label{Eqn:Pulse2Opt}
	\end{eqnarray}
\end{subequations}
The momentum quadratures have been entirely swapped, as desired, because the contribution of the initial mechanical position to $P_{M}^{\prime}$ due to free evolution has been coherently cancelled by the contribution of $X_{L}^{\prime}$ (which depends on the initial position) from the QND interaction. However, it is clear that $X_{M}^{\prime}$ ($X_{L}^{\prime}$) still retains some `memory' of the initial momentum $P_{M}$ ($P_{L}$) of the mechanical (optical) oscillator; this needs to be erased for the swap to be completed. Fortunately, this can be achieved by a second pair of $\pi/2$ rotations and a third QND interaction, \viz{}
\begin{subequations}
\label{Eqns:Pulse3}
	\begin{eqnarray}
	X_{M}^{\prime} = -P_{L}, & \; & P_{M}^{\prime} = X_{L}, \label{Eqn:Pulse3Mech}\\
	X_{L}^{\prime} = -P_{M}, & \; & P_{L}^{\prime} = X_{M}.  \label{Eqn:Pulse3Opt}
	\end{eqnarray}
\end{subequations}
This is a perfect state swap, up to local rotations (which we shall not consider), because the mechanical system now carries only fluctuations from the initial optical state, and vice versa.

\begin{figure}[t!]
\centering
\def\svgwidth{0.5\columnwidth}
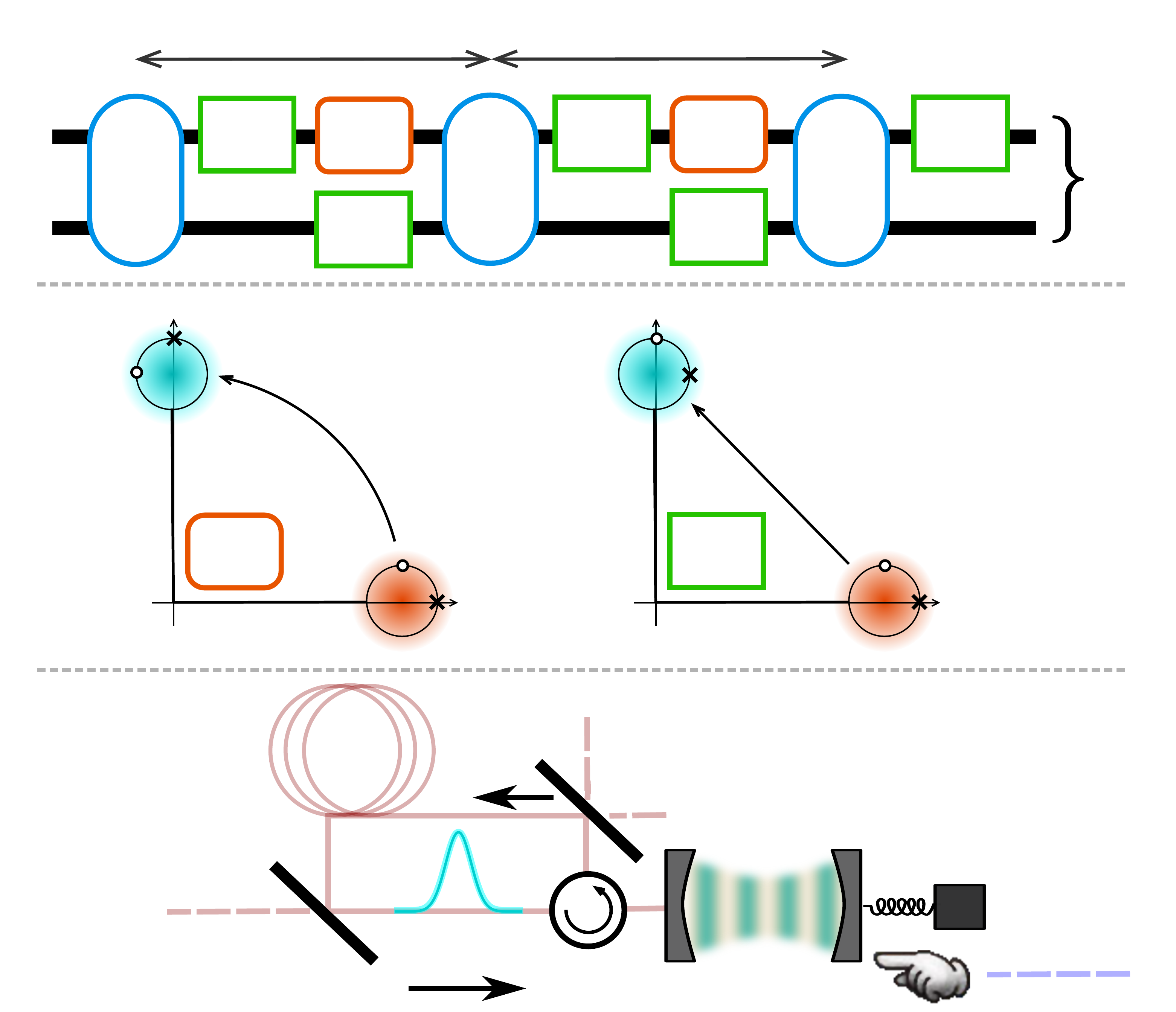
\caption{\label{Fig:Schema} Schematic of our pulsed optomechanical interface.\\
\textit{i}) Quantum circuit representation (operations proceeding from left to right). The input density operator is separable, $\hat{\rho} = \hat{\rho}_{M}\otimes\hat{\rho}_{L}$. An open-loop mechanical displacement ($D_{M}$) negates the classical momentum kick imparted by the mean amplitude of the pulse during the quantum non-demolition (QND) interaction. $R_{M}$ denotes a $\pi/2$ mechanical rotation (under free evolution) and $D_{L}$ is the matching optical displacement (noise rotation). Immediately before the second QND interaction the state of each oscillator is correlated with both inputs ($\hat{\rho}_{M}$ \& $\hat{\rho}_{L}$). In the ideal case the final QND interaction separates the output into the form $\hat{\rho} = \hat{\rho}_{L}\otimes\hat{\rho}_{M}$, but in general the output state need not be separable. \\
\textit{ii}) Illustration of the effect of $R_{M}$ and $D_{L}$ on Gaussian states. Each `ball \& stick' represents a noise ellipse of the state's Wigner function.  $R_{M}$ rotates the entire state (shown with non-zero mean amplitude for clarity). $D_{L}$ is a displacement of the optical state; note, however, that it effectively rotates the noise quadratures relative to the mean amplitude. The initial amplitude quadrature is marked by a cross, and the initial phase quadrature by a circle.\\
\textit{iii}) A potential physical architecture for the interface. The pulse carrying the initial optical quadratures is injected at the switch (upper right, S) and circulates counter-clockwise. After a QND interaction with the optomechanical cavity (centre) it is coupled into a delay arm via the switch and held for a quarter of a mechanical period. The pulse then encounters a bright, temporally-matched coherent pulse at the asymmetric beamsplitter (ABS, lower left), which perform a displacement of the pulse (rotation of the optical noise). This process repeats. After the final QND interaction the switch is opened and the output pulse leaves the interface.}
\end{figure}

\section{Decoherence Mechanisms} \label{Sec:DecoherenceMechs}

The fundamental limits on the interface's performance are imposed by optical loss and thermomechanical noise.

Optical loss---such as coupling inefficiencies, scattering and absorption---is readily treated by coupling the circulating pulse to a vacuum bath on a beamsplitter with efficiency $\eta_{L}$ (unity for perfect operation). The resulting nonhomogeneous evolution is
\[
	X_{L} \rightarrow \sqrt{\eta_{L}}X_{L} + \sqrt{1-\eta_{L}}X_{V},
\]
where $X_{V}$ is vacuum noise. An analogous relation holds for $P_{L}$. The properties of the vacuum mode are $\expect{X_{V}^{2}} = \expect{P_{V}^{2}} = 1$ and $\expect{X_{V}P_{V}}_{S} = 0$. Here we have introduced the symmetrised expectation value, $\expect{X_{V}P_{V}}_{S} = \half \expect{X_{V}P_{V}+P_{V}X_{V}} = \Re{\expect{X_{V}P_{V}}}$.

Mechanical decoherence is induced by two intimately related mechanisms: loss of energy from the oscillator into a heat bath, and random excitation of the oscillator by the bath. These processes may be described by the Langevin equations for the damped evolution (\cf{} \cite{Benguria1981}),
\begin{subequations}
\label{Eqns:DampedLangevin}
	\begin{eqnarray}
		\dot{X}_{M} & = & +\omega_{M} P_{M} \label{Eqn:Xdot}\\
		\dot{P}_{M} & = & -\omega_{M}X_{M} + \sqrt{2\Gamma}\hat{\xi}-\Gamma P_{M}.
		\label{Eqn:Damping}
	\end{eqnarray}
\end{subequations}
Momentum-dependent damping occurs at a rate $\Gamma$ whilst excitations enter via the random force $\hat{\xi}$ at a rate of $\Gamma\times\Br{\n_{B}+1/2}$, where $\n_{B}$ is the equilibrium occupancy of the oscillator, related to the bath temperature $T$ through the Bose-Einstein distribution (hence $\n_{B} \approx k_{B}T/\hbar\omega_{M}$ for large temperatures).

Non-zero mechanical damping reduces the mechanical resonance frequency to $\omega_{M}\sigma$, where $\sigma = \Br{1-\Gamma^{2}/4\omega_{M}^{2}}^{1/2} \leq 1$. It will be convenient to introduce the parameter $\epsilon = \Gamma/2\sigma\omega_{M}$, which is small in the limit of high mechanical quality ($Q_M \approx \omega_{M}/\Gamma \gg 1$). We may solve \eqref{Eqns:DampedLangevin} over a quarter cycle of evolution (Appendix~\ref{App:FreeEvoln}), giving
\begin{eqnarray*}
	X_{M} & \rightarrow & \sqrt{\eta_{M}}\Sq{+\epsilon X_{M}+\frac{1}{\sigma}P_{M}+\Delta X_{M}} \\
	P_{M} & \rightarrow & \sqrt{\eta_{M}}\Sq{-\epsilon P_{M}-\frac{1}{\sigma}X_{M}+\Delta P_{M}}
\end{eqnarray*}
where we have introduced noise increments $\Delta X_{M}$ and $\Delta P_{M}$ which depend on time integrals of $\hat{\xi}$ (given explicitly in Appendix~\ref{App:FreeEvoln}). For a high-quality oscillator $\eta_{M}$ is the amount by which an initial coherent excitation is attenuated over a time of one quarter period (so for an initial excitation amplitude $\beta$, $\mod{\beta}^{2} \rightarrow \eta_{M}\mod{\beta}^{2}$ over one quarter cycle); hence, by analogy with $\eta_{L}$, we call this quantity the mechanical efficiency. It is given by $\eta_{M} = \exp\Cu{-\pi\epsilon}$.

We close this section with some remarks about achieving $\chi = -1$ in the face of experimental imperfections and decoherence. Firstly, we wish to emphasise that many recent pulsed optomechanics experiments have entered a closely-related regime where the optomechanical cooperativity (see \eqref{Eqn:Coop}) exceeds $\n_{B}$ (\eg{} \cite{Palomaki2013b,Andrews2015}). The essential difference is that our scheme simultaneously requires $\kappa \sim g_{0}\sqrt{N}$ to be satisfied, implying that $g_{0}\sqrt{N} \gg \omega_{M}$. This necessitates the use of a very large number of pulse photons, because we are by assumption operating with $g_{0} \ll \omega_{M}$ and $\omega_{M} \ll \kappa$. Such high-$N$ pulses could be readily generated by acousto-optic or electro-optic (\eg{} \cite{Vanner2013}) modulation of a pump beam over the timescales relevant to our scheme. For example, a pulse lasting $1\%$ of a mechanical period at $\omega_{M}/2\pi = 100.2$~kHz requires an average power of $\sim 10$~mW to reach $N\sim 10^{9}$ (\cf{} example experimental parameters given in Table~\ref{Table:ExperimentalParams}). The chief technical issue expected to be encountered in this large $N$ limit is heating due to optical absorption, which is often a problem in continuous-wave optomechanics experiments. We present example calculations in Appendix~\ref{App:ChiNegOne} that show, using real experimental parameters, that such heating remains manageable for our pulsed scheme even in the regime of $\chi = -1$.

\begin{table}[t!]
\caption{\label{Table:ExperimentalParams} Exemplary parameters for use with the three-pulse protocol, taken from existing experiments. The mechanical frequency $\omega_{M}$, zero-point coupling rate $g_{0}$ and optical linewidth $\kappa$ are taken from silicon nitride microstring resonators evanescently coupled to silica microsphere whispering gallery modes \cite{Brawley2016}; the decay rate $\Gamma$ of comparable oscillators has previously been observed to be as low $2\pi \times 31$~mHz \cite{Brawley2012,Khosla2013}. With these specifications and a bath temperature of $4$~K the oscillator may be ground state cooled ($\expect{\hat{n}_{M}^{\prime}} = 0.606$). Reducing the bath temperature to $50$~mK (\eg{} \cite{Meenehan2014}) would permit $\expect{\hat{n}_{M}^{\prime}} = 0.008$ and preparation of the $\ket{1}$ motional state with infidelity $1-\mathcal{F} = 2.3\%$.}
\centering
	\begin{tabular}{c c | c c }
	$\omega_{M}/2\pi$ & $100.2$~kHz & 		$\Gamma/2\pi$ & $31$~mHz \\
	$\kappa/2\pi$ & $25.6$~MHz & 			$g_{0}/2\pi $ & $75$~Hz \\
	$N$ & $7.28\times 10^{9}$ &				$C$ & $7.72 \times 10^{5}$ \\
	$T$ & $\left\{\begin{tabular}{c} $4$~K \\ $50$~mK \end{tabular}\right.$ &
		$\n_{B}$ & $\left\{\begin{tabular}{c} $8.32\times 10^{5}$ \\ $1.04\times 10^{4}$ \end{tabular}\right.$
	\end{tabular}
\end{table}

\section{Extended Model} \label{Sec:ExModel}

The three-pulse protocol introduced above may easily be generalised to the case in which the pulses have unequal interaction strengths, which we shall denote by $\upBr{\chi}{j}$ where $j = 1,2,3$ labels each pulse. This could be arranged by using the control (displacement) pulses to alter the mean number of envelope photons $\upBr{N}{j}$ between QND interactions.

It will prove convenient to introduce a set of `pulse strengths' $\upBr{\mu}{j}$ which take into account the level of decoherence present; these are given by
\begin{eqnarray*}
\upBr{\mu}{1} & = & -\upBr{\chi}{1}\upBr{\mu}{2}, \\
\upBr{\mu}{2} & = & -\sqrt{\eta_{L}\eta_{M}} \upBr{\chi}{2}/\sigma, \\
\upBr{\mu}{3} & = & -\upBr{\chi}{3}\upBr{\mu}{2}.
\end{eqnarray*}
Note that in the absence of decoherence the perfect state swap is achieved when $\upBr{\mu}{j} = 1 \; \forall \; j$.

The general transformation between initial and final quadratures is a nonhomogeneous linear map,
\begin{equation}
	\bm{X}^{\prime} = M\bm{X} + \bm{F} \label{Eqn:NonHomoTransform}
\end{equation}
in which the vector of quadrature operators $\bm{X} = \trans{\Br{X_{M}, P_{M},X_{L}, P_{L}}}$ is acted upon by the matrix $M$ and there is a vector of zero-mean noise contributions $\bm{F}$ which depends upon $\upBr{\mu}{2}$, $\upBr{\mu}{3}$ and the noise increments $\Delta X_{M}$, $\Delta P_{M}$, $X_{V}$ and $P_{V}$   (see Appendix~\ref{App:AllPulse}).

The explicit matrix form of $M$ is
\begin{equation}
	\left(
	\begin{array}{c c c c}
	\mu^{\Br{1}}-\eta_{M} & 0 & 0 & -\mu^{\Br{2}} \\
	\epsilon\sigma\Sq{\mu^{\Br{3}}-\mu^{\Br{1}}} & \mu^{\Br{3}}-\eta_{M} &  M_{2,3} & \mu^{\Br{2}} \epsilon\sigma \\
	-\mu^{\Br{2}} \epsilon\sigma & -\mu^{\Br{2}} & \mu^{\Br{1}}-\eta_{L} & 0 \\
	M_{3,1} &  0 & 0 & \mu^{\Br{3}}-\eta_{L}
	\end{array}
	\right),
	\label{Eqn:M}
\end{equation}
in which
\begin{eqnarray*}
	M_{2,3} & = & \frac{1}{\mu^{\Br{2}}}\Sq{\eta_{L}\mu^{\Br{3}}+\mu^{\Br{1}}\Br{\eta_{M}-\mu^{\Br{3}}}}, \\
	M_{3,1} & = &\frac{1}{\mu^{\Br{2}}}\Sq{\eta_{M}\mu^{\Br{3}}+\mu^{\Br{1}}\Br{\eta_{L}-\mu^{\Br{3}}}}.
\end{eqnarray*}

To achieve a perfect swap the anti-diagonal elements of $M$ are desired to be $\pm 1$, while the diagonal blocks---which determine how complete the swap is---should be zero.

In what follows we will restrict ourselves to the case in which the optical and mechanical modes are initially separable and zero-mean ($\expect{\bm{X}} = \bm{0}$). Note that the latter may always be arranged by local operations on the input states, and that the mean amplitudes are unchanged by the interface.

\section{Ground State Cooling} \label{Sec:Cooling}

Ground state cooling is a convenient starting point for several schemes that aim to generate non-classical states of motion \cite{Chen2013, Kleckner2008}, fundamental tests of quantum mechanics and many quantum information processing protocols (\eg{} see \cite{Liu2015}). We shall refer to systems with average occupancies less than one ($\expect{\hat{n}_{M}^{\prime}} < 1$) as `ground state cooled' or `near ground state cooled'. Examples include \cite{OConnell2010, Teufel2011, Chan2011}.

Clearly the three-pulse state swap protocol introduced here can achieve optomechanical cooling in the bad-cavity limit; a coherent optical pulse carries only vacuum (zero temperature) noise, which can be swapped onto the mechanical system. In order to formulate a ground state cooling criterion 
we 
calculate the evolution of the mechanics--light covariance matrix (\cf{} Appendix~\ref{App:GroundCooling}).

The mechanical occupancy after a three--pulse state swap $\expect{\hat{n}_{M}^{\prime}}$ (given explicitly in \eqref{eqn:NumOutGeneric}) may be minimised with respect to the three pulse strengths following the procedure outlined in Appendix~\ref{App:GroundCooling}. In the high-$Q_{M}$ limit the minimum is achieved at $\upBr{\mu}{j} \approx 1$, yielding
\begin{equation}
	\expect{\hat{n}_{M}^{\prime}}_{min}  \approx  \frac{\pi\epsilon}{4}\Cu{3\Br{2\n_{B}+1}-2\eta_{L}} +\frac{1}{4}\Cu{\frac{1}{\eta_{L}}-1}.  \label{Eqn:NumOut}
\end{equation}

If the decoherence is dominated by thermomechanical noise, which is the case in the regime of interest (assuming $\n_{B} \gg 1$), we may insert \eqref{Eqn:NumOut} into the ground state criterion $\expect{\hat{n}_{M}^{\prime}} < 1$ to show that ground state cooling is possible only when the additional criterion
\begin{equation}
\n_{B} < \frac{4\omega_{M}}{3\pi\Gamma}
\label{Eqn:UnresolvedCriterionGamma}
\end{equation}
is satisfied. As discussed in \S~\ref{Sec:DecoherenceMechs} (\cf{} Table~\ref{Table:ExperimentalParams}), meeting the requirement $\upBr{\mu}{j}\approx 1$ is experimentally tractable. The further condition \eqref{Eqn:UnresolvedCriterionGamma} is essentially a requirement of quantum coherent oscillation (\ie{} less than one phonon entering the oscillator per mechanical period). Many experimental systems boast $Q_{M}$ large enough to satisfy this at temperatures achievable with conventional refrigeration technologies (\eg{} \cite{Brawley2012}). Recent devices may permit this regime to even be reached at room temperature ($Q_{M} \sim 10^{8}$ at $\omega_{M}/2\pi \sim 150$~kHz) \cite{Norte2016}.

Other strategies for optomechanical cooling, such as resolved-sideband cooling or cold damping, typically express their criteria for ground state cooling in terms of the optomechanical cooperativity $C$. We will therefore recast \eqref{Eqn:UnresolvedCriterionGamma} into a requirement on the cooperativity.

$C$ is a dimensionless parameter describes how strongly the quantum back-action of the optomechanical coupling (due to amplitude noise) perturbs the dynamics of the mechanical oscillator: in the steady state
\[
	C = \frac{4g_{0}^{2}\mod{\alpha}^{2}}{\Gamma\kappa} \; \mathrm{(steady \; state)}.
\]
A natural quantity to use in the pulsed setting is the instantaneous cooperativity time-averaged over the entire three-pulse sequence; thus
\begin{eqnarray}
	C & = & \frac{4 g_{0}^{2}}{\Gamma\kappa}\times \frac{\sigma\omega_{M}}{\pi} \intfin{t}{0}{\pi/\sigma\omega_{M}} \mod{\alpha\td}^{2} \label{Eqn:Coop} \\
	& = & \frac{1}{8\pi\epsilon}\Br{ \frac{\ghost{\upBr{\mu}{1}}^{2} + \ghost{\upBr{\mu}{3}}^{2}}{\ghost{\upBr{\mu}{2}}^{2}} + \frac{\sigma^{2} \ghost{\upBr{\mu}{2}}^{2} }{\eta_{L}\eta_{M}} } \nonumber
\end{eqnarray}
where in the first line $\alpha\td$ includes contributions from all three pulses.

\begin{figure}[tb]
\centering
\def\svgwidth{0.5\columnwidth}
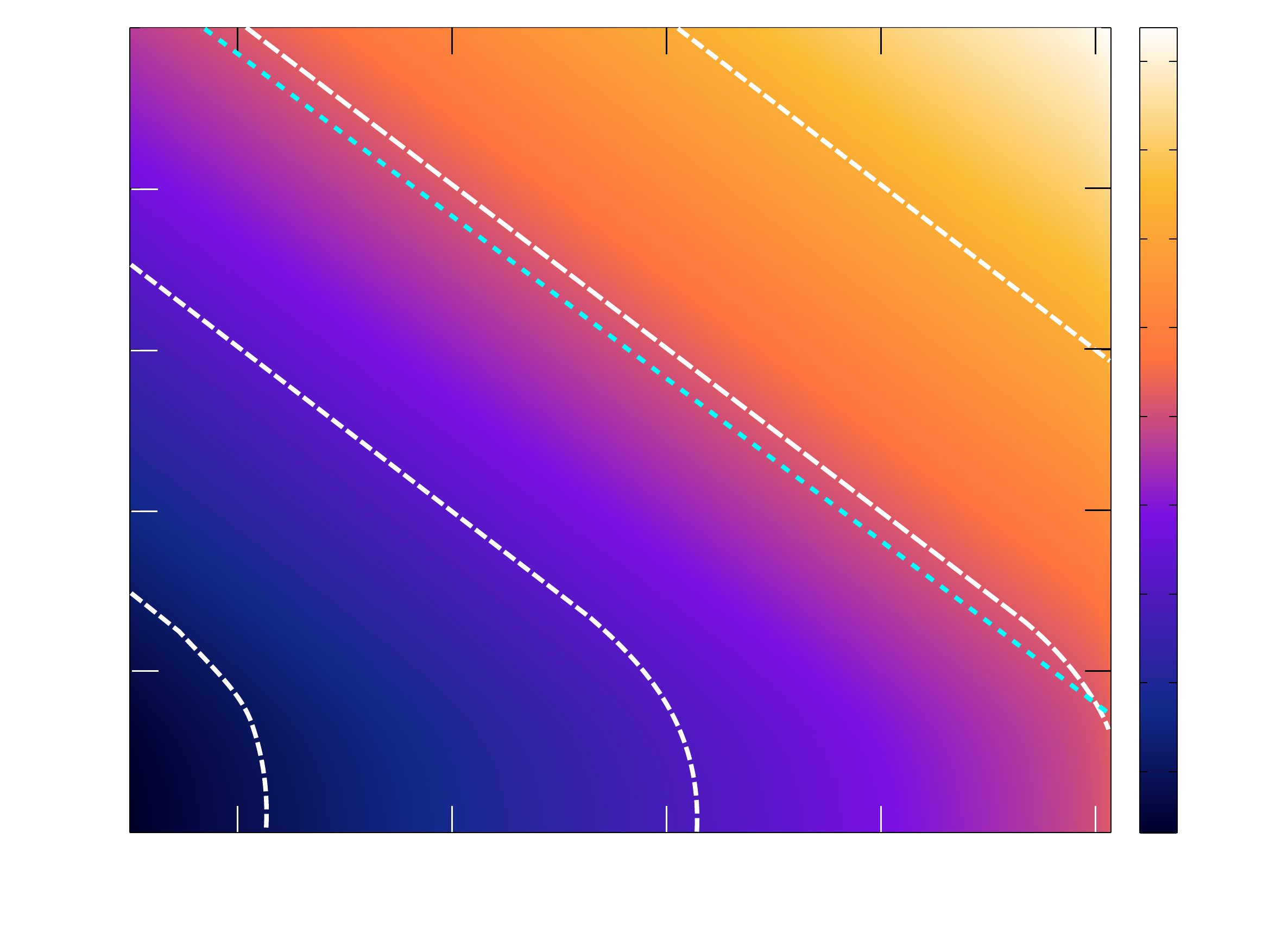
\caption{\label{Fig:CoolingSurf} Final occupancy of the mechanical oscillator after a complete three-pulse cooling sequence as a function of damping rate $\Gamma$ and bath occupancy $\n_{B}$. White contours show lines of equal (normalised) heating rates ($\Gamma \Br{\n_{B}+1/2}/\omega_{M}$); the shoulder at low $n_{B}$ is where vacuum fluctuations begin to dominate over thermal noise. The cyan dashed line demarcates the zone where ground state cooling ($\expect{\hat{n}_{M}^{\prime}} \leq 1$) is possible, as given by \eqref{Eqn:UnresolvedCriterion}. We have assumed $\upBr{\mu}{j} = 1$.}
\end{figure}

Substituting this definition into \eqref{Eqn:UnresolvedCriterionGamma} and noting that $\upBr{\mu}{j} \approx 1 \; \forall \; j$ for optimum cooling yields
\begin{equation}
\n_{B} < \frac{16}{3\Br{2+\eta_{L}^{-1}}} C \leq \Br{\frac{4}{3}}^{2} C.
\label{Eqn:UnresolvedCriterion}
\end{equation}
Thus, although operating in a pulsed manner and in a fundamentally different regime, our interface's ground state cooling criterion is remarkably similar to that of resolved-sideband cooling ($\n_{B}<C$), the current `gold standard' for optomechanical cooling \cite{Marquardt2007}. We note also that it also resembles the criterion for feedback cooling to near the ground state in the bad-cavity regime, namely $\n_{B} < 8 C$ \cite{Bennett2014}. However, it must be remembered that in \eqref{Eqn:UnresolvedCriterion} the cooperativity must be evaluated with $\upBr{\mu}{j} \approx 1$ because a near-perfect state swap is only possible in this regime: merely having a large cooperativity does not guarantee near ground state cooling using the three-pulse protocol.

In the case that the pulse strengths are not equal to unity we find that significant cooling is still possible so long as the deviation in $\mu$ is less than a threshold value which scales inversely with $\sqrt{Q_{M}}$ (Appendix~\ref{App:Tolerance}, Fig.~\ref{Fig:CoolingCross}); high-$Q_{M}$ oscillators are therefore more sensitive to imperfections in $\mu$.

The validity of \eqref{Eqn:UnresolvedCriterion} is confirmed by numerical calculations of the final phonon occupancy for a variety of $\upBr{\mu}{j}$, $\Gamma$ and $\n_{B}$, as shown in Fig.~\ref{Fig:CoolingSurf} (and Fig.~\ref{Fig:CoolingCross}). In all figures we have set $\eta_{L} = 1$.

Fig.~\ref{Fig:CoolingSurf} shows the expected trend of decreasing final occupancy as the oscillator $Q_{M}$ is increased and the initial temperature is decreased. The analytical threshold for near-ground-state cooling (\eqref{Eqn:UnresolvedCriterion}) is in excellent agreement with the calculations. Note that the lines of equal heating rate are curved because there is a non-zero incoming noise contribution from the mechanical bath even when the ambient temperature is zero.

Near ground state cooling can be realised using the three--pulse protocol with achievable experimental parameters, as shown in Table~\ref{Table:ExperimentalParams}. This cooling will persist for a time proportional to $\Br{\Gamma\n_{B}}^{-1}$ after the conclusion of the protocol. We note that these same parameters also permit non-classical state transfer, as discussed below (\S~\ref{Sec:FockState}).

\section{Fock State Preparation} \label{Sec:FockState}

For more complicated quantum states the covariance matrix ceases to be a full description, so we now move to considering the evolution of the system's Wigner function. The Wigner function is a quasiprobability distribution (Appendix~\ref{App:Non-Gaussian}) which is permitted to be negative over parts of phase space; negativity is considered to be a `smoking gun' for non-classical states \cite{Kleckner2008}.

\begin{figure}[t!]
\centering
\def\svgwidth{0.5\columnwidth}
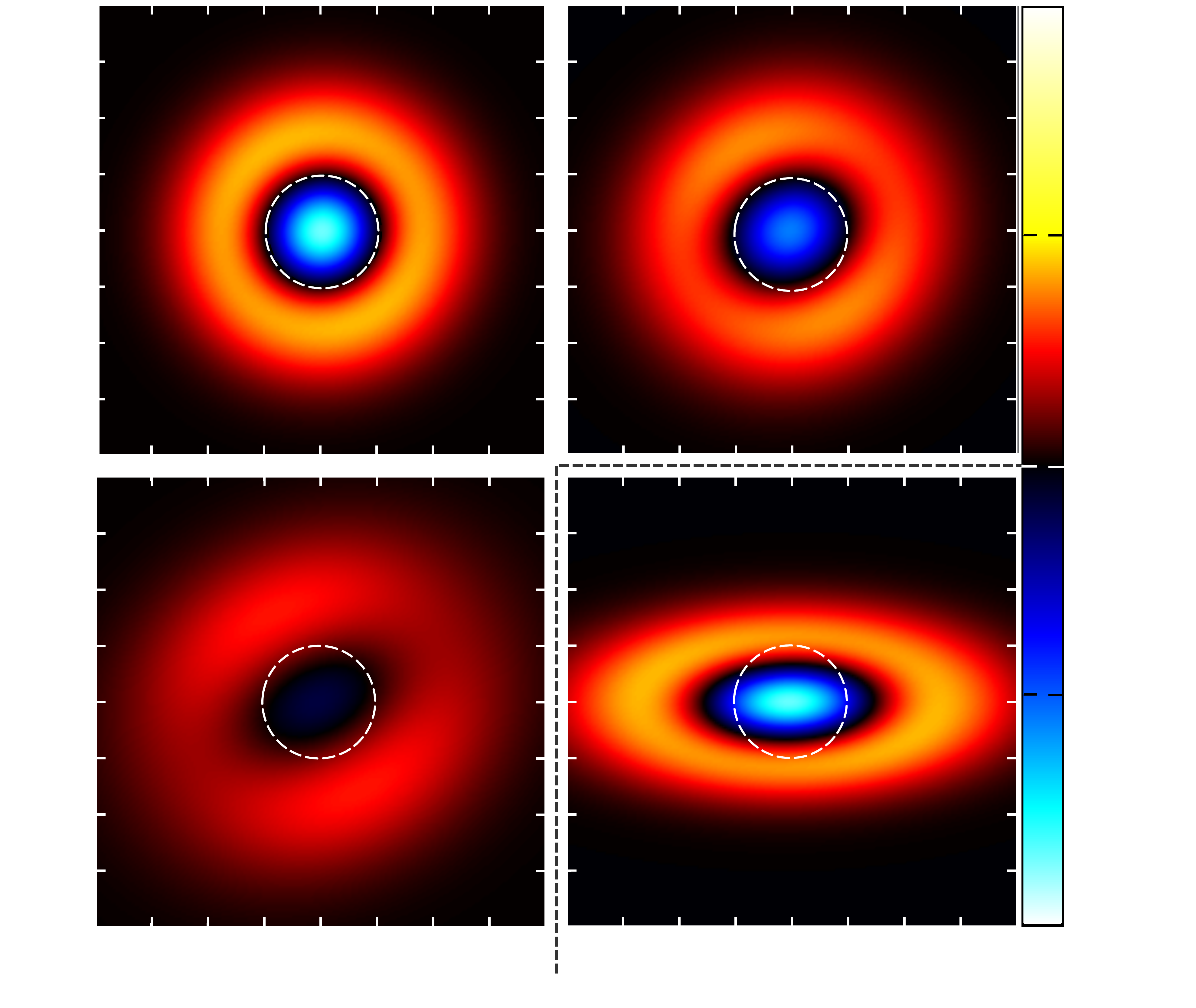
\caption{\label{Fig:Decoherence} \textit{i}), \textit{ii}) \& \textit{iii}): calculated output mechanical Wigner functions after transfer of an optical Fock state $\ket{1}$ in the presence of decoherence ($Q_{M} = \omega_{M}/\Gamma$) with $\upBr{\mu}{j} = 1$ for all pulses. These parameters are experimentally feasible. The mechanical state is initially thermal with $10$ phonons on average, simulating the result of a cooling pulse sequence, and the bath occupancy is fixed at $\n_{B} = 5\times 10^{4}$. Blue tones denote negative regions of the Wigner function and positive regions are red, passing through black at zero. The unit circle in white is the uncertainty contour of a vacuum state, shown for reference. Note that as the decoherence increases the output state becomes rotationally asymmetric and the Wigner negativity washes out. Due to the large optical bandwidth used in this scheme, it may be possible to transfer single photons generated using heralded spontaneous parametric down-conversion, which are naturally broad-band.\\
\textit{iv}) A `\Schrod{} kitten' state generated by squeezed transfer of a $\ket{1}$ state onto the mechanical oscillator. The bath conditions are the same as above, and $\upBr{\mu}{1}=\upBr{\mu}{3} = 1$. The target odd cat state amplitude is $\alpha$.}
\end{figure}

Our interface uses only linear interactions and therefore cannot generate Wigner negativity. It can, however, \textit{transfer} negativity from one system to another. We demonstrate this numerically by considering the preparation of a single phonon Fock state (Appendix~\ref{App:Fock}). Fock states have definite energy but completely undefined phase, and as such are highly nonclassical. As may be seen in Fig.~\ref{Fig:Decoherence} (panels \textit{i}--\textit{iii}), transfer of a single-photon state onto the mechanical oscillator is possible with realistic mechanical parameters (\cf{} Table~\ref{Table:ExperimentalParams}). Note that the state becomes rotationally asymmetric when the noise entering from the thermal bath is significant, and that increasing the heating rate leads to degradation of the Wigner negativity. This is consistent with thermalisation of the system during the protocol.

Higher-number mechanical Fock states may also be prepared. We show the infidelity (1-$\mathcal{F}$, where $\mathcal{F}$ is the Schumacher fidelity \cite{Schumacher1996}) of this operation in Fig.~\ref{Fig:Infidelity} (panel \textit{i}). Note that higher-$n$ Fock states are more difficult to prepare because of their intrinsically larger susceptibility to environmental noise.

\section{Squeezed Transfer} \label{Sec:Squeezed}

The interface may be made to perform additional non-trivial operations on the states involved. Here we consider fixing $\upBr{\mu}{1,3}= 1$ and allowing $\upBr{\mu}{2}$ to take on non-unit values. In this case \eqref{Eqn:M} becomes (setting $\eta_{L} = \eta_{M} = 1$)
\begin{equation}
	M\Br{\upBr{\mu}{2}} = \left(
	\begin{array}{c c c c}
	0 & 0 & 0 & -\mu^{\Br{2}} \\
	0 & 0 &  1/\upBr{\mu}{2} & 0 \\
	0 & -\mu^{\Br{2}} & 0 & 0 \\
	1/\upBr{\mu}{2} &  0 & 0 & 0
	\end{array}
	\right),
	\label{Eqn:Msqueezed}
\end{equation}
which may be decomposed into a state transfer, \ala{} \eqref{Eqns:Pulse3}, followed by local squeezing operations (one on each output) with equal squeezing parameters $\xi = -\ln{\upBr{\mu}{2}}$. Thus $\upBr{\mu}{2} < 1$ yields position-squeezed output states, and momentum-squeezed outputs may be prepared with $\upBr{\mu}{2} > 1$.

In principle, it is then straightforward to generate unconditionally squeezed motion by transferring the noise of a coherent pulse onto the oscillator with $\upBr{\mu}{2} \neq 1$.

A more intriguing application of this squeezed transfer is generation of small-amplitude mechanical \Schrod{} cat states (coherent state superpositions), termed `kitten' states. As has been demonstrated in optical systems, these may be `bred' into large-amplitude cat states by a number of schemes \cite{Etesse2015, Laghaout2013, Takahashi2008}; this could result in a macroscopic superposition of a massive object. Creating such a state is a research goal of foundational interest \cite{Chen2013}, and is also sufficient to allow universal quantum computation \cite{Jeong2002, Ralph2003} with phonons.

Any state of the form
\[
\ket{\Psi^{\Br{-}}} = \frac{\ket{\alpha} - \ket{-\alpha}}{\sqrt{2\Br{1-\e^{-2\mod{\alpha}^{2}}}}},
\]
is an odd cat state. Taking $\alpha$ to be small and real, we have
\begin{eqnarray*}
	\ket{\upBr{\Psi}{-}} & = & \frac{\ket{\alpha}-\ket{-\alpha}}{\sqrt{2\Br{1-\e^{-2\mod{\alpha}^{2}}}}}\\
	& = & \sqrt{\frac{2}{\e^{+\mod{\alpha}^{2}}-\e^{-\mod{\alpha}^{2}}}}\sum_{j \in \mathrm{odd}}^{\infty} \frac{\alpha^{j}}{\sqrt{j!}}\ket{j} \\
	& \approx & \frac{\alpha}{\mod{\alpha}}\Br{\ket{1} + \frac{\alpha^{2}}{\sqrt{6}}\ket{3} + \mathcal{O} \Br{\alpha^{4}}}.
\end{eqnarray*}
Note that this contains only odd Fock state contributions.

\begin{figure}[t!]
\centering
\def\svgwidth{0.5\columnwidth}
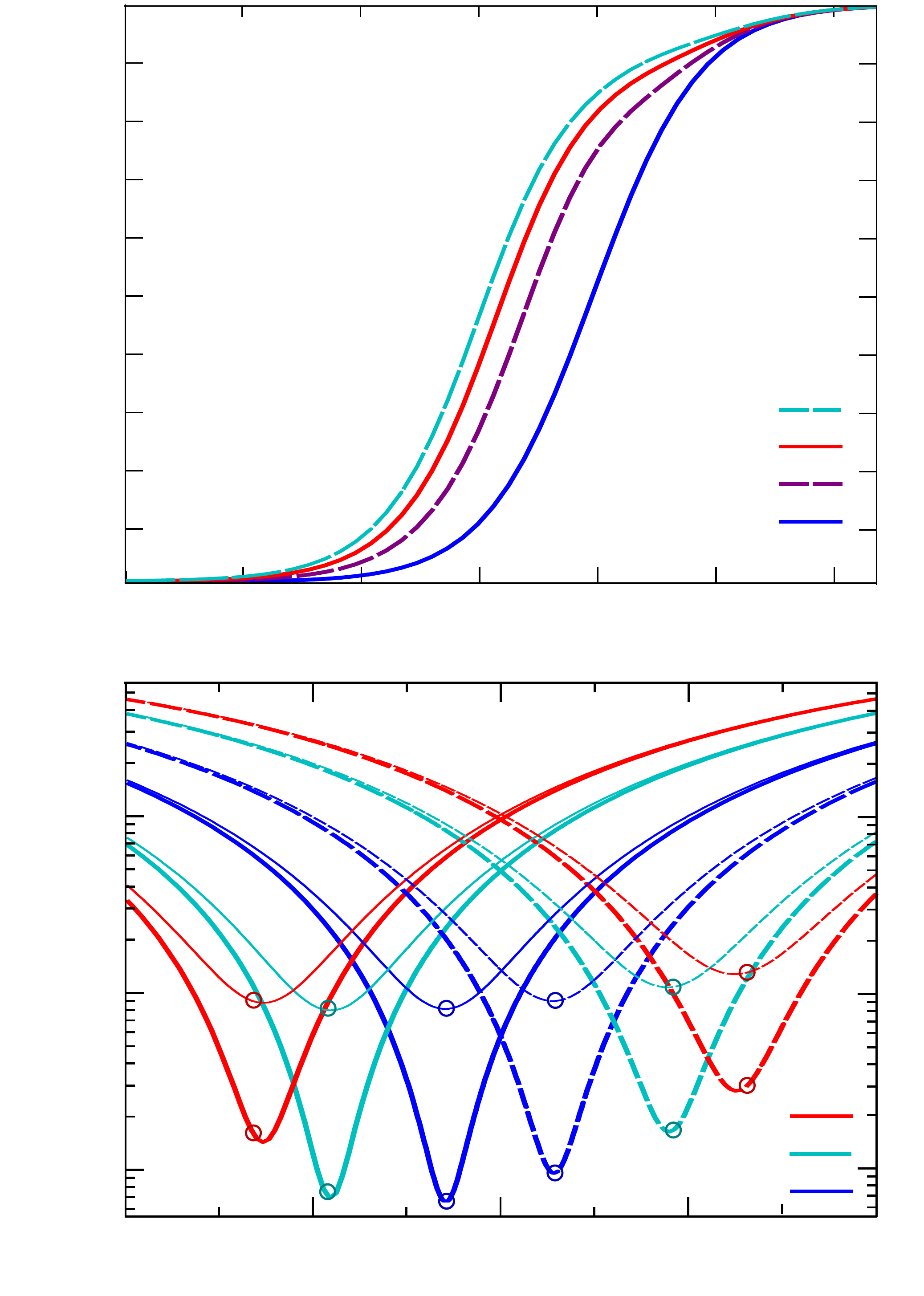
\caption{\label{Fig:Infidelity}
\textit{i}) The infidelity of mechanical Fock state preparation via pulsed optomechanical state transfer. The target states are $\ket{n}$ as indicated. All pulse strengths are fixed at unity and parameters are experimentally realisable. The initial mechanical state in all cases is thermal, containing $10$ phonons (with a bath of $\n_{B} = 5\times 10^{4}$).\\
\textit{ii}) Calculation of the infidelity of mechanical kitten state preparation using pulsed optomechanical state transfer with squeezing. Each colour corresponds to a given mean number of phonons in the target kitten, as indicated, with dashes corresponding to target states with imaginary $\alpha$ and solid lines indicating $\alpha \in \R$. Thick lines show the decoherence-free case ($\Gamma/\omega_{M} = 0$, initial mechanical state $\ket{0}$), and results with the small but achievable ratio $\Gamma/\omega_{M} = 2.24\times 10^{-7}$ are shown as thin lines (initially thermal with 10 phonons, $\n_{B} = 5\times 10^{3}$). $\xi=-\ln \upBr{\mu}{2}$ is the magnitude of the squeezing induced by the transfer (with $\upBr{\mu}{1}=\upBr{\mu}{3}=1$). Circles indicate the \naive{} analytical estimates for optimum overlap with each target kitten state, $\xi = -\alpha^{2}/3$. The asymmetry is due to thermal noise being mixed into the output state differently for different values of $\upBr{\mu}{2}$.}
\end{figure}

A squeezer generates correlated pairs of photons; it follows that weakly squeezing $\ket{1}$ could give an approximation to an odd kitten state (\cf{} \cite{Lund2004}). For small squeezing parameters $\xi$,
\begin{eqnarray*}
	S\Br{\xi}\ket{1} & = & \exp\Cu{\half\Br{\xi^{*}\baop{a}^{2}-\xi\ghost{\bcop{a}}^{2}}}\ket{1} \\
	& = & \sum_{j=0}^{\infty}\Br{\half}^{j}\frac{\Br{\xi^{*}\baop{a}^{2}-\xi\ghost{\bcop{a}}^{2}}^{j}}{j!}\ket{1} \\
	& \rightarrow & \ket{1} - \xi\sqrt{\frac{3}{2}}\ket{3} + \mathcal{O} \Br{\xi^{2}}.
\end{eqnarray*}
By comparing these we see that the approximate kitten state has an amplitude given by $\xi = -\frac{\alpha^{2}}{3}$. Thus we obtain $\upBr{\mu}{2}=\exp\Cu{\alpha^{2}/3}$.

The Wigner function of such a state is given in Appendix~\ref{App:CatStateWigner} and shown with mild decoherence in Fig.~\ref{Fig:Decoherence} (panel \textit{iv}). It is important to note that the Wigner function contains two positive lobes along the $X_{M}^{\prime}$ axis (`alive' and `dead') separated by a region of quantum interference exhibiting Wigner negativity, proving that coherence between the lobes persists \cite{Kleckner2008}. Again, note that these parameters are experimentally realisable (although requiring very large $N$ to achieve the squeezing shown in Fig.~\ref{Fig:Decoherence}, panel \textit{iv}).

The degree to which the decohering squeezed single phonon state approximates a true cat state is shown in Fig.~\ref{Fig:Infidelity} (panel \textit{ii}). It is evident that large cat states cannot be produced directly in this manner, as the minimum infidelity rises monotonically with $\mod{\alpha}$; this is not due to imperfections in our interface, but from the approximate equality of an odd cat state and a squeezed single phonon state. Environmental noise clearly degrades the preparation process, both through decreasing the purity of the initial mechanical state and through destroying the phonon pair correlations required for squeezing. Even so, it is possible to push the infidelity to low values with realistic parameters (\cf{} Table~\ref{Table:ExperimentalParams}). We also note that other non-Gaussian quantum states of light can be transferred to the mechanics by our interface thus providing several routes for motional state engineering.

\section{Conclusion} \label{Sec:Conclusion}

We have shown how open--loop rotations and displacements plus repeated QND interactions may be used to construct an optomechanical interface which operates in the high-bandwidth ($\kappa \gg \omega_{M}$) limit of optomechanics, a natural operating regime for many devices. Our calculations predict that such an interface is within reach of current experimental parameters (\cf{} Table \ref{Table:ExperimentalParams}). Furthermore, we have shown how our three-pulse protocol may be extended to performing more complicated operations, including optical--mechanical state transfer with squeezing of the final modes; we note that it is also possible to perform a partial state swap which entangles the final modes, though we have not modelled that process here.

These results show that optomechanical systems in the unresolved sideband limit are capable of performing an optomechanical state-swap, and may prove important in the construction of quantum networks and studies of macroscopic quantum mechanics. Finally, we note that this three-pulse protocol could directly be applied in other circumstances where QND interactions with light are available, such as in spin ensembles.

\section*{Funding Information}
This work was funded by the Australian Research Council through the Centre of Excellence for Engineered Quantum Systems (EQuS, CE110001013), a Discovery Project grant (M.~R.~V., DP140101638) and a Future Fellowship (W.~P.~B., FT140100650); L.~S.~M. further acknowledges support from the Carlsberg Foundation.

\section*{Acknowledgments}

The authors thank Nicholas Wyatt, Andreas N\ae{}sby, Yauhen `Eugene' Sachkou and George A. Brawley for useful discussions.

\appendix
\bigskip
\begin{center}
\textbf{SUPPLEMENTARY INFORMATION}
\end{center}

We begin by describing the definition of the pulse fluctuations (Appendix~\ref{App:FluctuationsDef}) and modelling the loss processes which fundamentally limit the performance of our proposed three-pulse protocol (Appendix~\ref{App:Model}). This is followed by a derivation of the transformation between input and output quadratures in the presence of loss (Appendix~\ref{App:Assembling}). We outline the ground state cooling calculations in Appendix~\ref{App:GroundCooling}. Finally, we introduce the Wigner function and characteristic function description of our system, and give explicit forms of Gaussian and non-Gaussian (Fock and cat state) Wigner functions considered in the main manuscript (Appendix~\ref{App:Non-Gaussian}).

\section{Fluctuation Operators} \label{App:FluctuationsDef}

The optomechanical coupling Hamiltonian takes the form given in the main text (\cf{} \cite{Milburn2011, Aspelmeyer2013review});
\[
	\hat{H} = \hbar g_{0} X_{M}\bcop{a}\baop{a}.
\]
If we apply a bright pulse to the cavity (on resonance) then we may linearise around the intracavity field's mean amplitude $\alpha\td$. Doing so amounts to letting $\baop{a} \rightarrow \alpha\td + \delta\baop{a}\td$ where the displaced annihilation operator $\delta\baop{a}\td$ is zero mean. Then
\[
	\hat{H} = \hbar g_{0} X_{M} \Br{\mod{\alpha\td}^{2} + \alpha^{*}\delta\baop{a}\td + \alpha\td\delta\bcop{a}\td + \delta\bcop{a}\td\delta\baop{a}\td}.
\]
As is common, we neglect the `small' term $\delta\bcop{a}\td\delta\baop{a}\td$ because the optomechanical coupling is taken to be weak. Thus we arrive at
\[
	\hat{H} = \hbar \mod{\alpha\td} g_{0} X_{M} \bar{X}_{L}\td + \hbar g_{0} X_{M} \mod{\alpha\td}^{2},
\]
as given in the main text. From this we see that the amplitude fluctuations are defined by
\[
	\bar{X}_{L}\td = \frac{\alpha\td \delta\bcop{a}\td + \alpha^{*}\td\baop{a}\td }{\mod{\alpha\td}}.
\]
This reduces to the expression given in the main text if $\alpha$ is real and positive.

The conjugate fluctuations, representing phase noise of the field, may be written
\[
	\bar{P}_{L}\td = \I \frac{\alpha\td \delta\bcop{a}\td - \alpha^{*}\td\baop{a}\td }{\mod{\alpha\td}}.
\]
These definitions ensure that the amplitude and phase fluctuations transform correctly when the optical state is displaced. Specifically, if the state is displaced such that $\alpha \rightarrow \I \alpha$ then the optical noise is rotated by $\pi/2$. The utility of this definition will become apparent in Appendix~\ref{App:PulseDisplacement}.

\section{Loss Processes} \label{App:Model}

\subsection{Free Mechanical Evolution (Rotation)} \label{App:FreeEvoln}

For a linearly damped oscillator
\begin{eqnarray*}
\dot{X}_{M} & = & \omega_{M}P_{M}, \\
\dot{P}_{M} & = & -\omega_{M}X_{M} + \sqrt{2\Gamma}\hat{\xi}-\Gamma P_{M}.
\end{eqnarray*}
The noise operator $\hat{\xi}$ has the properties \cite{Benguria1981, GardinerZoller} 
\begin{eqnarray*}
\expect{\hat{\xi}\td\hat{\xi}\Br{t^{\prime}}} & = & \intfin{\omega}{0}{\infty} \Cu{\frac{\omega}{\pi\omega_{M}}\coth\Br{\frac{\hbar\omega}{2k_{B}T_{B}}} \cos{\omega\Br{t-t^{\prime}}}}\\
& & {} + \frac{\I}{2\omega_{M}}\Br{\frac{\partial}{\partial t}-\frac{\partial}{\partial t^{\prime}}}\delta\Br{t-t^{\prime}}, \\
\comm{\xi\td}{X_{M}\td} & = & \I\sqrt{2\Gamma}, \\
\comm{Y\td}{\xi\Br{t^{\prime}}} & = & \frac{1}{\omega_{M}}\sqrt{\frac{\Gamma}{2}} \de{}{t^{\prime}}\Cu{u\Br{t-t^{\prime}} \comm{Y\td}{X_{M}\Br{t^{\prime}}}},
\end{eqnarray*}
where $u$ is the piecewise step function equal to $1/2$ at the origin, $1$ for positive arguments and $0$ otherwise. $Y\td$ is an arbitrary system operator. These properties are necessary for preservation of the canonical commutator of $X_{M}$ and $P_{M}$ \cite{Benguria1981}. In the high-temperature limit (where $\hbar \omega \ll 2 k_{B} T_{B}$ for all relevant frequencies $\omega$) we may approximate the former by
\[
\expect{\hat{\xi}\td\hat{\xi}\Br{t^{\prime}}} = \Br{2\n_{B}+1}\delta\Br{t-t^{\prime}} + \frac{\I}{2\omega_{M}}\Br{\frac{\partial}{\partial t}-\frac{\partial}{\partial t^{\prime}}}\delta\Br{t-t^{\prime}}
\]
where $\n_{B}$ is the average occupancy of the mechanical mode when in thermal equilibrium with its bath.

The equations of motion may then be integrated to yield
\begin{eqnarray*}
	X_{M}\td & = & \e^{-\Gamma t/2}\Sq{\cos\Br{\omega_{M}\sigma t}+\epsilon\sin\Br{\omega_{M}\sigma t}}X_{M}\Br{0} + \\
	& & \e^{-\Gamma t/2}\Sq{\frac{1}{\sigma}\sin\Br{\omega_{M}\sigma t} P_{M}\Br{0} + \Delta X_{M}\td}, \\
	P_{M}\td & = & \e^{-\Gamma t/2}\Sq{\cos\Br{\omega_{M}\sigma t}-\epsilon\sin\Br{\omega_{M}\sigma t}}P_{M}\Br{0} + \\
	& & \e^{-\Gamma t/2}\Sq{\frac{-1}{\sigma}\sin\Br{\omega_{M}\sigma t} X_{M}\Br{0} + \Delta P_{M}\td},
\end{eqnarray*}
where the damping rate $\Gamma$ induces an offset to the oscillation frequency, characterised by
\begin{eqnarray*}
	\sigma & = & \sqrt{1-\Br{\Gamma/2\omega_{M}}^{2}} \\
		& = & \Br{1+\epsilon^{2}}^{-1/2},
\end{eqnarray*}
with $\epsilon = \Gamma/2\omega_{M}\sigma$ (we assume that the oscillator is underdamped throughout). Noise entering from the thermal bath appears in these expressions as
\begin{eqnarray*}
	\Delta X_{M}\td & = & \sqrt{2\Gamma} \intfin{t^{\prime}}{0}{t} \e^{\Gamma t^{\prime}/2}\frac{1}{\sigma}\sin\Br{\omega_{M}\sigma\Br{t-t^{\prime}}} \hat{\xi}\Br{t^{\prime}},\\
	\Delta P_{M}\td & = & \sqrt{2\Gamma} \intfin{t^{\prime}}{0}{t} \e^{\Gamma t^{\prime}/2}\hat{\xi}\Br{t^{\prime}} \times\\
	& & \left[\cos\Br{\omega_{M}\sigma \Br{t-t^{\prime}}}-\epsilon\sin\Br{\omega_{M}\sigma\Br{t-t^{\prime}}}\right].
\end{eqnarray*}

When evaluated after a quarter of a mechanical cycle ($X_{M}$ and $P_{M}$ in the following expressions are implicitly evaluated at time zero unless otherwise stated, whilst the noise operators are evaluated at the end of the quarter cycle)
\begin{eqnarray*}
	X_{M}\td & = & \sqrt{\eta_{M}}\Sq{\epsilon X_{M}+\frac{1}{\sigma}P_{M}+\Delta X_{M}},\\
	P_{M}\td & = & \sqrt{\eta_{M}}\Sq{-\epsilon P_{M}-\frac{1}{\sigma}X_{M}+\Delta P_{M}},
\end{eqnarray*}
and the following commutators apply;
\begin{eqnarray*}
	\comm{\Delta X_{M}}{\Delta P_{M}} & = & 2\I\Br{\eta_{M}^{-1}-1}, \\
	\comm{X_{M}}{\Delta X_{M}} & = & \comm{X_{M}}{\Delta P_{M}} = 0, \\
	\comm{P_{M}}{\Delta X_{M}} & = & + 2\I \epsilon, \\
	\comm{P_{M}}{\Delta P_{M}} & = & - 2\I \sigma\epsilon^{2}.
\end{eqnarray*}

The efficiency $\eta_{M} = \exp\Cu{-\pi\epsilon}$ is approximately the amount by which a coherent excitation of the oscillator decays over the course of one quarter mechanical period: for an initial coherent excitation amplitude $\beta$ we have $\mod{\beta}^{2} \rightarrow \eta_{M} \Br{1/\sigma^{2}+\epsilon^{2}}\mod{\beta}^{2} = \eta_{M} \Br{1+2\epsilon^{2}}\mod{\beta}^{2}$, which is approximately $\eta_{M}\mod{\beta}^{2}$ for high-$Q_{M}$ oscillators.

\subsection{Pulse Absorption} \label{App:ChiNegOne}

Here we show that our scheme does not suffer from significant heating due to absorption (see \S~\ref{Sec:DecoherenceMechs}) by considering an explicit potential realisation. We model a microstring mechanical resonator of thickness $t_{R} = 54$~nm, width  $w_{R} = 10$~$\upmu$m and length $\ell_{R} = 1$~mm; such devices may be fabricated from silicon nitride (\cf{} Table~\ref{Table:ExperimentalParams}) \cite{Brawley2016}. Note that all of the parameters below have been realised in existing experiments.

Using the formula for the absorption of a dielectric plate given by \cite{Wilson2012(thesis)} (Eqn~(3.7) thereof) and the measured imaginary refractive index of SiN \cite{Chakram2014} we estimate that the fraction of incident power absorbed by the oscillator is $f \sim 10^{-5}$. The energy deposited into the resonator by a single pulse is therefore $E \approx \hbar\omega_{p}f N 2 \mathcal{F}/\pi$, where $\mathcal{F}$ is the cavity finesse (\cf{} \cite{Vogell2013}) and $\omega_{p}$ is the pulse's central frequency. We have taken the energy of a pulse photon to be $\hbar \omega_{p}$, ignoring its spectral width.

It takes a characteristic time for thermal energy deposited at the centre of the string to leave, given by $\omega_{th}^{-1} = \rho_{R}c_{R}\ell_{R}^{2}/\kappa_{th}$, where $\rho_{R}$ is the device's mass density, $c_{R}$ is its specific heat capacity and $\kappa_{th}$ is its thermal conductivity. We take $\rho_{R} = 3.18\times 10^{3}$~kg~m$^{-3}$, $c_{R} = 711$~J~kg$^{-1}$~K$^{-1}$ and $\kappa_{th} = 150$~W~m$^{-1}$~K$^{-1}$ (values for Si$_{3}$N$_{4}$ taken from \cite{CRCHandbookMechanical}, tables 40, 88 and 104). Given the geometry described here, $\omega_{M} \gg \omega_{th} \approx 260$~Hz. This means that we may treat the dynamics of the heat distribution as very slow compared to the mechanical motion of interest.

During the full protocol time of half a mechanical period, thermal energy deposited at the centre of the string has propagated a distance (root-mean-squared) of $L = \sqrt{2\pi\kappa_{th}/\rho_{R}c_{R}\omega_{M}}$. Thus the volume of heated material may be approximated as $V_{R} \approx t_{R} w_{R} L$. The resulting change in temperature of the material over this volume is $\delta T = E/\pi \rho_{R} c_{R} V_{R}$. As a consequence, the occupancy of the bath seen by the mechanical mode changes by (at worst)
\begin{eqnarray*}
	\delta \n_{B} & = & 3 k_{B}\delta T/\hbar\omega_{M} \\
	& = & 3\frac{2 f k_{B}\omega_{p}\mathcal{F}}{\pi^{2} \omega_{M} \rho_{R} c_{R} V_{R}} N.
\end{eqnarray*}
The overall factor of three accounts for the fact that there are three pulses.

If we employ a pulse with photons at a central wavelength of $1559$~nm \cite{Brawley2016} we obtain $\delta \n_{B} \approx 10^{-9}\times \mathcal{F} N$. Thus for a finesse of $100$ and photon number as in Table~\ref{Table:ExperimentalParams} the change in the effective bath occupancy is approximately $10\%$ at $50$~mK.

This result is in many ways a worst-case scenario. Firstly, the change in temperature seen by the mode of interest is likely to be lower than calculated here because the mode extends over a larger region of the string. Secondly, alternative resonator geometries (which are not long and thin) can permit much lower $\delta \n_{B}$ because of their larger volume and faster thermal diffusion times. We are therefore confident that heating will not (in principle) prevent one from achieving $\chi = -1$.

\subsection{Optical Displacements and Loss} \label{App:PulseDisplacement}

A displacement of the optical field in any direction not parallel to its initial phase angle has the effect of rotating the quadrature fluctuations (because the instantaneous fluctuations have been defined to be parallel and perpendicular to the mean amplitude, \cf{} Appendix~\ref{App:FluctuationsDef}). For instance, consider the special case of a displacement sending the mean amplitude $\alpha$ to $\I \alpha$; this yields the transformation
\begin{eqnarray*}
X_{L} & \rightarrow & P_{L}, \\
P_{L} & \rightarrow & -X_{L},
\end{eqnarray*}
which is precisely a $\pi/2$ \textit{rotation} of the noise (by contrast, a rotation of the optical field leaves the noise unchanged, because the fluctuations are also rotated).

Such an operation may be realised by interfering the circulating pulse with a very bright incoming pulse, temporally matched to the former but offset in phase, on a highly asymmetric beamsplitter. Given that the noise associated with this operation may be made arbitrarily small in principle (by selecting an extreme asymmetry and applying a more powerful displacement pulse) we will neglect it, or else absorb it into the total efficiency $\eta_{L}$ defined below.

Any optical losses due to inefficiencies within the recirculating loop may be partially compensated by altering the magnitude and phase angle of the displacement pulse, but contamination of the fluctuation quadratures by vacuum noise cannot be reversed. Losses may be modelled as an asymmetric beamsplitter which transmits $\sqrt{\eta_{L}}$ of the circulating pulse ($0 < \eta_{L} \leq 1$) and couples in vacuum noise proportional to $\sqrt{1-\eta_{L}}$.

\begin{eqnarray*}
	X_{L} & \rightarrow & \sqrt{\eta_{L}}X_{L}+\sqrt{1-\eta_{L}}X_{V}, \\
	P_{L} & \rightarrow & \sqrt{\eta_{L}}P_{L}+\sqrt{1-\eta_{L}}P_{V}.
\end{eqnarray*}

\section{Modelling the Dynamics} \label{App:Assembling}

\subsection{QND Interaction} \label{App:QND}

The QND interaction (including open loop displacement of the mechanics) may be written as
\[
	\bm{X} \rightarrow M_{QND}^{\Br{j}}\bm{X}
\]
with
\[
M_{QND}^{\Br{j}} = \Br{
\begin{array}{c c c c}
1 & 0 & 0 & 0 \\
0 & 1 & \chi^{\Br{j}} & 0 \\
0 & 0 & 1 & 0 \\
\chi^{\Br{j}} & 0 & 0 & 1
\end{array}}.
\]

\subsection{Evolution During Pulse--Delay--Rotate Cycle} \label{App:SinglePulse}

We now consider the net evolution induced by a QND interaction followed by a quarter-cycle delay and an optical displacement. The full protocol consists of two of these cycles followed by a further QND interaction.

The evolution induced by a single pulse--displace--delay cycle may be expressed as a nonhomogeneous linear equation
\begin{equation*}
	\bm{X} \rightarrow M^{\Br{j}}\bm{X} + \bm{F}^{\Br{j}}
\end{equation*}
where $M^{\Br{j}}$ is a square matrix characterised by the interaction strength $\chi^{\Br{j}} = \frac{-8g_{0}\sqrt{N^{\Br{j}}}}{\kappa}$ and the efficiencies $\eta_{L}$ and $\eta_{M}$;
\[
M^{\Br{j}} = \Br{
\begin{array}{c c c c}
\epsilon\sqrt{\eta_{M}} & \sqrt{\eta_{M}}/\sigma & \sqrt{\eta_{M}}\chi^{\Br{j}}/\sigma & 0 \\
-\sqrt{\eta_{M}}/\sigma & -\epsilon \sqrt{\eta_{M}} & -\epsilon\chi^{\Br{j}} \sqrt{\eta_{M}} & 0 \\
\sqrt{\eta_{L}}\chi^{\Br{j}} & 0 & 0 & \sqrt{\eta_{L}} \\
0 & 0 & -\sqrt{\eta_{L}} & 0 
\end{array}}.
\]
The inhomogeneous term corresponds to thermal and vacuum noises; $\bm{F}^{\Br{j}} = \bm{f}^{\Br{j}}+\Br{1-\eta_{L}}\bm{v}^{\Br{j}}$ where
\[
\bm{f}^{\Br{j}} = \sqrt{\eta_{M}}\trans{\Br{\Delta X_{M}^{\Br{j}}, \; \Delta P_{M}^{\Br{j}},\; 0,\; 0}}
\]
and
\[
\bm{v}^{\Br{j}} = \trans{\Br{0,\; 0,\; X_{V}^{\Br{j}}, \; P_{V}^{\Br{j}}}}
\]
in which $X_{V}^{\Br{j}}$ ($P_{V}^{\Br{j}}$) are vacuum noise operators and $\Delta P_{M}^{\Br{j}}$ and $\Delta X_{M}^{\Br{j}}$ are evaluated at $t = \pi/2\omega_{M}\sigma$ (\ie{} at one-quarter mechanical period).

The noise has the following covariance, with other terms vanishing;

\begin{eqnarray*}
\expect{\Delta X_{M}^{\Br{j}} \Delta X_{M}^{\Br{k}}} & = & \expect{\Delta P_{M}^{\Br{j}} \Delta P_{M}^{\Br{k}}}\\
& = & \delta_{j,k}\Br{2\n_{B}+1}\Br{\frac{1}{\eta_{M}}-1-2\epsilon^{2}},\\
\expect{\Delta X_{M}^{\Br{j}} \Delta P_{M}^{\Br{k}}}_{S} & = & 4\delta_{j,k}\Br{2\n_{B}+1}\frac{\epsilon}{\sigma}, \\
\expect{X_{V}^{\Br{j}} X_{V}^{\Br{k}}} & = & \expect{P_{V}^{\Br{j}} P_{V}^{\Br{k}}}\\ 
& = & \delta_{j,k}, \\
\expect{X_{V}^{\Br{j}} P_{V}^{\Br{k}}}_{S} & = & 0.
\end{eqnarray*}
$\delta_{j,k}$ is the Kronecker delta.

\subsection{Full Evolution} \label{App:AllPulse}

Taking the evolution given above and iterating (remembering that the final pulsed interaction is not followed by a rotation \& optical displacement) yields
\[
	\bm{X}^{\prime} = M\bm{X}+\bm{F}
\]
where
\begin{eqnarray*}
	M & = & \upBr{M_{QND}}{3}\upBr{M}{2}\upBr{M}{1}, \\
	\bm{F} & = & \upBr{M_{QND}}{3}\Sq{\upBr{M}{2}\upBr{\bm{F}}{1} + \upBr{\bm{F}}{2}}.
\end{eqnarray*}

The explicit matrix form of $M$ is
\[
	\left(
	\begin{array}{c c c c}
	\mu^{\Br{1}}-\eta_{M} & 0 & 0 & -\mu^{\Br{2}} \\
	\epsilon\sigma\Sq{\mu^{\Br{3}}-\mu^{\Br{1}}} & \mu^{\Br{3}}-\eta_{M} &  M_{2,3} & \mu^{\Br{2}} \epsilon\sigma \\
	-\mu^{\Br{2}} \epsilon\sigma & -\mu^{\Br{2}} & \mu^{\Br{1}}-\eta_{L} & 0 \\
	M_{3,1} &  0 & 0 & \mu^{\Br{3}}-\eta_{L}
	\end{array}
	\right),
\]
in which
\begin{eqnarray*}
	M_{2,3} & = & \frac{1}{\mu^{\Br{2}}}\Sq{\eta_{L}\mu^{\Br{3}}+\mu^{\Br{1}}\Br{\eta_{M}-\mu^{\Br{3}}}}, \\
	M_{3,1} & = &\frac{1}{\mu^{\Br{2}}}\Sq{\eta_{M}\mu^{\Br{3}}+\mu^{\Br{1}}\Br{\eta_{L}-\mu^{\Br{3}}}}.
\end{eqnarray*}

It is instructive to consider the form of $M$ in some ideal cases.

\subsubsection{State Swap} \label{App:StateSwap}

Setting $\upBr{\mu}{j}=1$ with no decoherence ($\epsilon = 0$, $\eta_{M} = \eta_{L} = 1$) yields
\[
	M = \left(
	\begin{array}{c c c c}
	0 & 0 & 0 & -1 \\
	0 & 0 &  1 & 0\\
	0 & -1 & 0 & 0 \\
	1 &  0 & 0 & 0
	\end{array}
	\right).
\]

\subsubsection{Squeezed State Swap} \label{App:SqueezedStateSwap}

Setting $\upBr{\mu}{1,3}=1$ with no decoherence gives
\[
	M = \left(
	\begin{array}{c c c c}
	0 & 0 & 0 & -\upBr{\mu}{2} \\
	0 & 0 &  1/\upBr{\mu}{2} & 0\\
	0 & -\upBr{\mu}{2} & 0 & 0 \\
	1/\upBr{\mu}{2} &  0 & 0 & 0
	\end{array}
	\right).
\]
This may be factorised into a state swap (as given above) followed by the application of the two local squeezing operations
\begin{eqnarray*}
	S & = & \exp\Cu{\half\xi^{*}\ghost{\baop{a}}^{2}-\half\xi\ghost{\bcop{a}}^{2}}\exp\Cu{\half\xi^{*}\ghost{\baop{b}}^{2}-\half\xi\ghost{\bcop{b}}^{2}} \\
	& \rightarrow & \left(
	\begin{array}{c c c c}
	\upBr{\mu}{2} & 0 & 0 & 0 \\
	0 & 1/\upBr{\mu}{2} & 0 & 0\\
	0 & 0 & \upBr{\mu}{2} & 0 \\
	0 &  0 & 0 & 1/\upBr{\mu}{2}
	\end{array}
	\right),
\end{eqnarray*}
which have equal squeezing parameters $\xi = -\ln\upBr{\mu}{2}$.

\section{Ground State Cooling Calculations} \label{App:GroundCooling}

Under the assumptions listed in the main text (\S~ \ref{Sec:Cooling}), plus the assumption that the noise is uncorrelated with the initial state, it is straightforward to show that the output covariance matrix
\[
	V\sb{\bm{X}\bm{X}}^{\prime} = \Re{\expect{\bm{X}^{\prime}\trans{\bm{X}^{\prime}}}}
\]
is related to the input by (\cf{} \eqref{Eqn:NonHomoTransform})
\[
	V\sb{\bm{X}\bm{X}}^{\prime} = MV\sb{\bm{X}\bm{X}}\trans{M}+V\sb{\bm{F}\bm{F}}.
\]
The first diagonal block of $V\sb{\bm{X}\bm{X}}^{\prime}$ corresponds to the output mechanical covariance matrix, the second represents the output optical state, and the antidiagonal blocks describe correlations between the two subsystems.

In the high-$Q_{M}$ limit we may expand this result about $\epsilon = 0$ and determine the final mechanical variances.
\begin{subequations}
\label{Eqns:VarApp}
	\begin{eqnarray}
		\expect{\ghost{X_{M}^{\prime}}^{2}} & = & \expect{X_{M}^{2}}\Br{\upBr{\mu}{1}-1}\Br{\upBr{\mu}{1}+2\pi\epsilon-1} \nonumber\\
			& & {}+\expect{P_{L}^{2}}\ghost{\upBr{\mu}{2}}^{2} + 2\pi\epsilon\Br{2\n_{B}+1} \nonumber\\
			& & {}+ \Br{\frac{1}{\eta_{L}}-1}\ghost{\upBr{\mu}{2}}^{2},
			\label{Eqn:VarXApp} \\
		\expect{\ghost{P_{M}^{\prime}}^{2}} & = & \expect{P_{M}^{2}}\Br{\upBr{\mu}{3}-1}\Br{\upBr{\mu}{3}+2\pi\epsilon-1} \nonumber\\
			& & {}+ \expect{X_{M}P_{M}}_{S} 2\epsilon \Br{\upBr{\mu}{3}-1}\Br{\upBr{\mu}{3}-\upBr{\mu}{1}} \nonumber\\
			& & {}+ \expect{X_{L}P_{L}}_{S} 2\epsilon \Br{\upBr{\mu}{1}+\upBr{\mu}{3}\eta_{L}-\upBr{\mu}{1}\upBr{\mu}{3}} \nonumber\\
			& & {}+ \expect{X_{L}^{2}}\ghost{\upBr{\mu}{2}}^{-2}\Cu{\upBr{\mu}{1}\Br{\upBr{\mu}{3}-1}-\eta_{L}\upBr{\mu}{3}} \nonumber\\
			& & {}\times \Cu{\upBr{\mu}{1}\Br{\upBr{\mu}{3}+2\pi\epsilon-1}-\eta_{L}\upBr{\mu}{3}} \nonumber\\
			& & {}+ \pi\epsilon\Br{2\n_{B}+1}\Cu{2-\upBr{\mu}{3}\Br{2-\upBr{\mu}{3}}} \nonumber\\
			& & {}+ \Br{\frac{\upBr{\mu}{3}}{\upBr{\mu}{2}}}^{2}\Br{1-\eta_{L}^{2}}.
			\label{Eqn:VarPApp}
	\end{eqnarray}
\end{subequations}

It is safe to assume that the initial mechanical state is thermal, \viz{}
\[
	\expect{X_{M}^{2}} = \expect{P_{M}^{2}} = 2\n_{B}+1,\; \expect{X_{M}P_{M}}_{S}=0,
\]
and because a coherent pulse carries only vacuum noise we have
\[
	\expect{X_{L}^{2}} = \expect{P_{L}^{2}} = 1,\; \expect{X_{L}P_{L}}_{S}=0.
\]
The final occupancy of the mechanical oscillator is then readily calculated using
\begin{equation}
	\expect{\hat{n}_{M}^{\prime}} = \frac{1}{4}\Sq{\expect{\ghost{X_{M}^{\prime}}^{2}}+\expect{\ghost{P_{M}^{\prime}}^{2}}-2}. \label{Eqn:NumDef}
\end{equation}
%
%
This yields
\begin{eqnarray}
	4\expect{\hat{n}_{M}^{\prime}} & = & \Br{2\n_{B}+1}\left[ \Br{\upBr{\mu}{1}-1}\Br{\upBr{\mu}{1}+2\pi\epsilon-1} \right. \label{eqn:NumOutGeneric} \\
	& & {} +\Br{\upBr{\mu}{3}-1}\Br{\upBr{\mu}{3}+2\pi\epsilon-1} \nonumber \\
	& & \left. {}+ \pi\epsilon\Br{4-\upBr{\mu}{3}\Br{2-\upBr{\mu}{3}}}\right] \nonumber \\
	& & {}+\frac{1}{\ghost{\upBr{\mu}{2}}^{2}}\left[\ghost{\upBr{\mu}{3}}^{2}-2\eta_{L}\upBr{\mu}{1}\upBr{\mu}{3}\Br{\upBr{\mu}{3}+\pi\epsilon-1} \right. \nonumber \\
	& & \left. {} + \ghost{\upBr{\mu}{1}}^{2}\Br{\upBr{\mu}{3}-1}\Br{\upBr{\mu}{3}+2\pi\epsilon-1}\right] \nonumber \\
	& & {}+\frac{\ghost{\upBr{\mu}{2}}^{2}}{\eta_{L}}-2. \nonumber
\end{eqnarray}

To minimise this, note that $\expect{\hat{n}_{M}^{\prime}}$ may be divided into a term proportional to $\Br{2 \n_{B}+1}$ and a term independent of the thermal bath. The former is dependent only upon $\upBr{\mu}{1}$ and $\upBr{\mu}{3}$, whilst the latter depends on all three strengths. We therefore choose to approximately minimise the occupancy by minimising the two terms separately.

Minimising the bath-dependent term yields, to first order in $\epsilon$,
\[
	\upBr{\mu}{1,3} \approx 1-\pi \epsilon \approx \eta_{M}.
\]
We then substitute these into the remaining noise contribution and select $\upBr{\mu}{2}$ accordingly. Thus for optimum cooling we require
\begin{eqnarray*}
	\upBr{\mu}{1} & \approx & \eta_{M} \\
	\upBr{\mu}{2} & \approx & \frac{1+\eta_{M}}{2} \eta_{L}^{1/4}\\
	\upBr{\mu}{3} & \approx & \eta_{M}. \\
\end{eqnarray*}
As $\eta_{M}$ \& $\eta_{L}$ approach $1$ we obtain $\upBr{\mu}{j} = 1$ to a very good approximation, as stated previously.

Evaluated at $\upBr{\mu}{j} = 1$,
\[
	\expect{\hat{n}_{M}^{\prime}}_{min} = \frac{\pi\epsilon}{4}\Cu{3\Br{2\n_{B}+1}-2\eta_{L}} +\frac{1}{4}\Cu{\frac{1}{\eta_{L}}-1}.
\]
This result is given as \eqref{Eqn:NumOut} in the main text.

Note that in order to ground state cool one must satisfy  $\eta_{L} > 1/5$ in addition to the criteria given in the main text (\eqref{Eqn:UnresolvedCriterion}).

\subsection{Tolerance to Pulse Strength Variations} \label{App:Tolerance}

\begin{figure}[hbt]
\centering
\def\svgwidth{0.5\columnwidth}
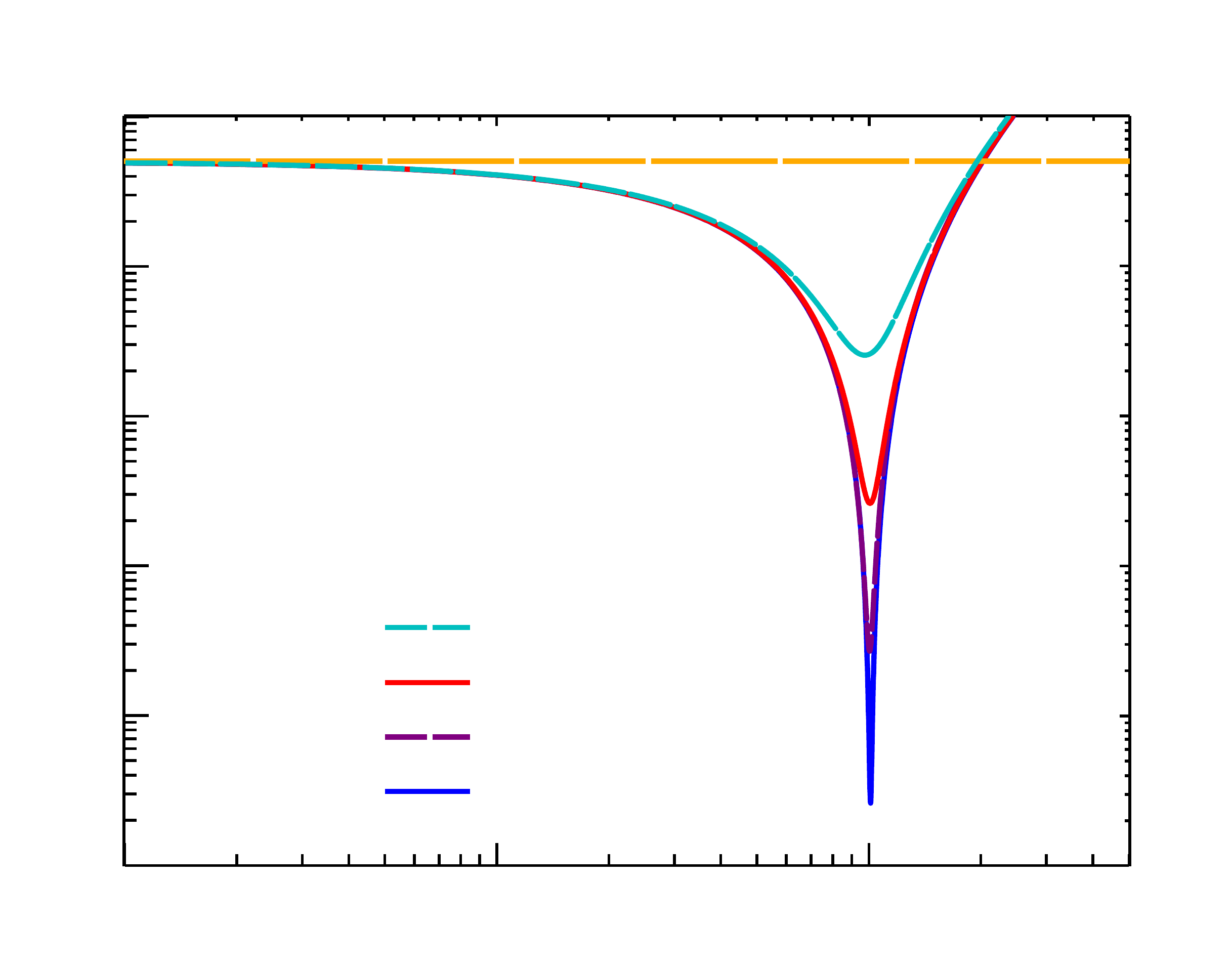
\caption{\label{Fig:CoolingCross} Final mechanical occupancy $\expect{\hat{n}_{M}^{\prime}}$ as a function of pulse strength $\upBr{\mu}{j} = \mu$ for different mechanical quality factors. The bath occupancy is fixed at $5000$, corresponding to a temperature of approximately $24$ mK for $\omega_{M}/ 2\pi = 100.2$~kHz (\cf{} Table~\ref{Table:ExperimentalParams}).}
\end{figure}

It is clear from Fig.~\ref{Fig:CoolingCross} that the system becomes less tolerant to variations in $\upBr{\mu}{j}$ for very high-$Q_{M}$ oscillators, which means that the mean number of photons in the coherent envelope of each pulse must be controlled to a very high degree when attempting to reach the mechanical ground state.

To quantify this we consider the case where all pulse strengths are equal (denoted by $\mu$ with no superscript). We first find the two values of $\mu$ for which
\[
	\expect{\hat{n}_{M}^{\prime}} = 2 \expect{\hat{n}_{M}^{\prime}}_{min}.
\]
The difference of these values is used as a proxy for the width of the cooling dip (\cf{} Fig.~\ref{Fig:CoolingCross}, main text). If $\n_{B} \gg 1$ and $\eta_{L} \gtrsim 0.1$ this reduces to
\[
	\mathrm{width} \approx \sqrt{6 \pi \epsilon}.
\]
Thus the tolerance of the protocol to imperfections in the pulse strengths scales as $\epsilon^{1/2}$. This is a very strong dependence in the high-$Q_{M}$ limit; for instance, a mechanical $Q_{M}$ of $10^{6}$ would require the value of $\mu$ to be stabilised to approximately one part in a thousand.

To translate this into terms of photon number, note that $\mathrm{width} \approx \Delta \mu \approx -\Delta \chi$, where $\Delta \mu$ ($\Delta\chi$) is the tolerance of the protocol to $\mu$ ($\chi$). Then we may expand the definition of $\chi$ to see that
\[
	\Delta \chi \approx \frac{-8g_{0}}{\kappa}\sqrt{N}\Br{\frac{\Delta N}{2N}},
\]
which becomes $\Delta \chi \approx -\Delta N/2N$ in the case that $\chi \approx -1$. Thus $\Delta N/N \approx 2\sqrt{6\pi \epsilon}$ is the tolerable fractional variation of the pulse photon number amplitude. For $Q_{M}$ of $10^{6}$ the fractional variation needs to be less than approximately $0.6\%$.

This $\sqrt{\epsilon}$ scaling may be explained by considering the two ways in which the thermal bath contributes to the final mechanical temperature. Firstly, there is noise entering the system during the protocol; this component is reasonably weakly dependent on $\mu$, and tends to set the minimum attainable final temperature. The second contribution is from the initial thermal state of the oscillator. The amount of this noise which is swapped out of the mechanical system depends strongly on $\mu$, and tends to set the shape of the $\expect{\hat{n}_{M}^{\prime}\Br{\mu}}$ curve (\cf{} Fig.~\ref{Fig:CoolingCross}, main text). If the former dominates then the strong $\mu$ dependence of the second term is washed out.

\section{Non-Gaussian States} \label{App:Non-Gaussian}

When considering non-Gaussian input states, such as Fock states, the covariance matrix is no longer an adequate description of the state. We therefore consider the Wigner function description of quantum states; for $n$ modes this is defined by
\begin{equation}
W\Br{\bm{r}} = \frac{1}{\Br{2\pi}^{2n}} \int \mathrm{d}^{2n}\bm{\beta} \; \Tr{\hat{\rho}\D{\bm{\beta}}}\exp\Cu{-\I \bm{r}\cdot \Omega \bm{\beta}}, \label{Eqn:WignerCharacteristicFunctionDefinition}
\end{equation}
in which the $2n\times 2n$ matrix $\Omega$ is the $n$-fold block diagonal matrix with diagonal blocks
\[
	\varpi = \Br{\begin{array}{c c} 0 & 1 \\ -1 & 0 \end{array}}
\]
and all other entries equal to zero (note that $\trans{\Omega} = - \Omega$ and $\Omega\trans{\Omega} = \trans{\Omega}\Omega = -\mathfrak{1}$, with $\mathfrak{1}$ being the identity matrix) \cite{WallsMilburn2008}. Points in phase space are denoted $\trans{\bm{r}} = \Br{x_{1}, p_{1}, x_{2}, \cdots}$ and we have suppressed the bounds of integration, which are at $\pm \infty$ for each quadrature. $\D{\bm{\beta}}$ is the multi-mode displacement operator with $\trans{\bm{\beta}} = \Br{\Re{\beta_{1}},\Im{\beta_{1}}, \Re{\beta_{2}},\cdots}$ giving the displacement of each mode.

The Wigner function is Fourier dual to the characteristic function $\chi\Br{\bm{\beta}} = \Tr{\hat{\rho}\D{\bm{\beta}}}$, \viz{}
\begin{equation}
	\chi\Br{\bm{\beta}} = \int \mathrm{d}^{2n}\bm{r} \; W\Br{\bm{r}} \exp\Cu{+\I\bm{r}\cdot\Omega\bm{\beta}}, \label{Eqn:CharacterDefine}
\end{equation}
and knowledge of either is sufficient to reconstruct the density operator $\hat{\rho}$ \cite{WallsMilburn2008}. It will prove useful to work in the reciprocal domain ($\bm{\beta}$) for the calculations which follow.

\subsection{Wigner Functions Used in This Work} \label{App:WignerFuncs}

\subsubsection{Gaussian States}

Gaussian Wigner functions have the following form, where $\delta\bm{r} = \bm{r} - \expect{\bm{X}} = \trans{\Br{x_{1}-\expect{X_{1}},\;p_{1}-\expect{P_{1}}, \cdots}}$:
\begin{subequations}
\label{Eqns:WignerCharacter}
	\begin{eqnarray}
	W\Br{\bm{r}} & = & \frac{1}{\Br{2\pi}^{n}\sqrt{\mod{V}}} \exp\Cu{-\half\trans{\delta\bm{r}} V^{-1}\delta\bm{r}}.\nonumber 
	\label{Eqn:GaussWigner} \\
	\chi\Br{\bm{\beta}} & = & \exp\Cu{-\half\trans{\bm{\beta}}\trans{\Omega} V \Omega \bm{\beta}}\e^{\I\expect{\bm{X}}\cdot \Omega\bm{\beta}}. \nonumber
	\label{Eqn:GaussCharacteristic}
	\end{eqnarray}
\end{subequations}

They are entirely specified by the mean values and $\expect{\bm{X}}$ and the covariance matrix $V = \Re{\expect{\bm{X}\trans{\bm{X}}}}$.

\subsubsection{Fock states} \label{App:Fock}

The Fock states $\ket{n}$ are non-Gaussian, having Wigner functions given by the Laguerre polynomials $\mathcal{L}\sb{n}$. We give these for a single mode only.
\begin{eqnarray*}
	\chi\sb{n}\Br{\beta} & = & \e^{-\half\mod{\beta}^{2}}\mathcal{L}\sb{n}\Br{\mod{\beta}^{2}},\\
	W\sb{n}\Br{\bm{r}} & = & \frac{1}{2\pi}\Br{-1}^{n}\exp\Cu{-\half\bm{r}\cdot\bm{r}}\mathcal{L}\sb{n}\Br{\bm{r}\cdot\bm{r}}.
\end{eqnarray*}

\subsubsection{Cat states} \label{App:CatStateWigner}

The final class of non-Gaussian state treated in this paper are the odd ($-$) cat states. For completeness we include the expressions for the even ($+$) cat states too. Given a finite amplitude $\alpha$ (for a single mode),
\begin{eqnarray*}
	\upBr{\chi\sb{\alpha}}{\pm}\Br{\beta} & = & \Br{1\pm \e^{-2\mod{\alpha}^{2}}}^{-1}\e^{-\half\mod{\beta}^{2}} \left[\cosh\Sq{\alpha^{*}\beta-\beta^{*}\alpha} \pm \e^{-2\mod{\alpha}^{2}}\cosh\Sq{\alpha^{*}\beta+\beta^{*}\alpha}\right],\\
	\upBr{W\sb{\alpha}}{\pm}\Br{\bm{r}} & = & \frac{\e^{-\half\bm{r}\cdot\bm{r}}}{2\pi\Br{1\pm \e^{-2\mod{\alpha}^{2}}}} \Sq{\e^{-2\mod{\alpha}^{2}}\cosh\Br{2\bm{r}\cdot\bm{\alpha}}\pm\cos\Br{2\bm{r}\cdot \varpi \bm{\alpha}}}.
\end{eqnarray*}

\subsection{Lossy Evolution} \label{App:LossyEvolution}

The output characteristic function may be written (according to its definition in the Heisenberg picture, for a linear system) as
\begin{eqnarray*}
	\chi^{\prime}\Br{\bm{\beta}} & = & \Tr{\hat{\rho} \exp\Sq{\I \bm{X}^{\prime}\cdot\Omega\bm{\beta}} }\\
	& = & \Tr{\hat{\rho} \exp\Sq{\I \Br{M\bm{X}+\bm{F}}\cdot\Omega\bm{\beta}}}\\
	& = & \Tr{\hat{\rho} \exp\Sq{\I\trans{\bm{\beta}}\trans{\Omega}M\bm{X}}\exp\Sq{\I\bm{F}\cdot\Omega\bm{\beta}}} \\
	& = & \Tr{\hat{\rho} \exp\Sq{\I\trans{\bm{\beta}}\trans{\Omega}M\bm{X}}}\Tr{\hat{\rho}\exp\Sq{\I\bm{F}\cdot\Omega\bm{\beta}}}.
\end{eqnarray*}
In going from line two to line three we used the \BCH{} lemma, and to arrive at line four we noted that expectation values of system and noise operators factorise under the approximations used here.

If we define a vector $\bm{\gamma} = \Omega \trans{M} \Omega \bm{\beta}$ then it is clear that the exponent of the first trace above becomes
\begin{eqnarray*}
	\trans{\bm{\beta}}\trans{\Omega}M\bm{X} & = & -\trans{\bm{\beta}}\trans{\Omega}M\Omega\trans{\Omega}\bm{X}\\
	& = & \trans{\bm{\beta}}\Br{-\trans{\Omega}M\Omega}\trans{\Omega}\bm{X}\\
	& = & \trans{\bm{\beta}}\Br{\Omega M\Omega}\trans{\Omega}\bm{X}\\
	& = & \trans{\bm{\gamma}}\trans{\Omega}\bm{X},
\end{eqnarray*}
whence
\[
	\chi^{\prime}\Br{\bm{\beta}} = \Tr{\hat{\rho} \exp\Sq{\I\bm{X}\cdot \Omega\bm{\gamma}}}\Tr{\hat{\rho}\exp\Sq{\I\bm{F}\cdot\Omega\bm{\beta}}}.
\]
We now recognise the term on the left as being simply the input characteristic function with its arguments rescaled to $\bm{\gamma}$, \viz{}
\[
	\chi\Br{\bm{\beta}} \rightarrow \chi\Br{\Omega \trans{M} \Omega \bm{\beta}}\Tr{\hat{\rho}\exp\Sq{\I\bm{F}\cdot\Omega\bm{\beta}}}.
\]

 Since we are already aware of the general form of a Gaussian characteristic function (\eqref{Eqn:GaussCharacteristic}) we may simply use this formula to find the kernel induced by Gaussian noise, \viz{}
\begin{eqnarray*}
	\chi_{F}\Br{\bm{\beta}}  & = & \Tr{\hat{\rho}\exp\Sq{\I\bm{F}\cdot\Omega\bm{\beta}}}\\
	& = & \exp\Cu{-\half\trans{\bm{\beta}}\trans{\Omega} V_{FF} \Omega \bm{\beta}}, \\
\end{eqnarray*}
where $V_{FF} = \Re{\expect{\bm{F}\trans{\bm{F}}}}$.

Let us now consider a two mode system. If the density matrix is separable (\ie{} $\hat{\rho} = \hat{\rho}_{M}\otimes\hat{\rho}_{L}$), as we have assumed, the input characteristic function factorises into the form
\[
	\chi\Br{\bm{\beta}} = \chi_{M}\Br{\bm{\beta}_{M}}\chi_{L}\Br{\bm{\beta}_{L}}
\]
where
\[
	\bm{\beta} = \Br{\begin{array}{c} \bm{\beta}_{M} \\ \bm{\beta}_{L} \end{array}}
\]

Let us define the matrices
\begin{eqnarray*}
	E_{M} = \Br{\begin{array}{cccc} 1 & 0 & 0 & 0 \\ 0 & 1 & 0 & 0 \end{array}}, \\
	E_{L} = \Br{\begin{array}{cccc} 0 & 0 & 1 & 0 \\ 0 & 0 & 0 & 1 \end{array}}.
\end{eqnarray*}
With these we may write $\bm{\beta}_{M,L} = E_{M,L}\bm{\beta}$. Thus, from the evolution above we can find that
\[
	\chi\Br{\Omega \trans{M}\Omega\bm{\beta}} = \chi_{M}\Br{E_{M}\Omega \trans{M}\Omega\bm{\beta}}\chi_{L}\Br{E_{L}\Omega \trans{M}\Omega\bm{\beta}}
\]

This allows us to write the transformation between input and output characteristic functions as
\[
	\chi_{M}\Br{\bm{\beta}}\chi_{L}\Br{\bm{\beta}} \rightarrow \chi_{M}\Br{E_{M}\Omega \trans{M}\Omega\bm{\beta}}\chi_{L}\Br{E_{L}\Omega \trans{M}\Omega\bm{\beta}} \chi_{F}\Br{\bm{\beta}}
\]

The final Wigner function of the mechanical oscillator is obtained by evaluating this at $\trans{\bm{\beta}} = \Br{\Re{\beta_{M}}, \Im{\beta_{M}}, 0, 0 }$ and applying the appropriate transformation (\eqref{Eqn:WignerCharacteristicFunctionDefinition}).

Note that this calculation is equivalent to determining the Wigner function by propagating the reduced density matrix of the mechanical mode in the \Schrod{} picture.

\subsection{Fidelity} \label{App:Fidelity}

The Schumacher fidelity may be recast as
\[
	\mathcal{F} = 4 \pi \int \mathrm{d}^{2}\bm{r} \; W^{\prime}\Br{\bm{r}}W_{target}\Br{\bm{r}}.
\]

\end{document}